\documentclass[sigconf]{aamas} 

\usepackage{balance} 
\usepackage{lipsum}
\usepackage{multicol,multirow}
\usepackage{array}
\usepackage{subcaption}
\usepackage{caption}
\usepackage{float}
\usepackage{algorithm}
\usepackage{algorithmicx}
\usepackage{algpseudocode}
\usepackage{eqparbox}
\usepackage{color}
\usepackage{xcolor}
\usepackage[verbose]{placeins}
\usepackage{enumitem}
\usepackage{amsmath}

\setcopyright{ifaamas}
\acmConference[AAMAS '24]{Proc.\@ of the 23rd International Conference
on Autonomous Agents and Multiagent Systems (AAMAS 2024)}{May 6 -- 10, 2024}
{Auckland, New Zealand}{N.~Alechina, V.~Dignum, M.~Dastani, J.S.~Sichman (eds.)}
\copyrightyear{2024}
\acmYear{2024}
\acmDOI{}
\acmPrice{}
\acmISBN{}

\acmSubmissionID{976}

\title[AAMAS-2024 Formatting Instructions]{Quantifying Agent Interaction in Multi-agent Reinforcement Learning for Cost-efficient Generalization}

\author{Yuxin Chen}
\affiliation{
  \institution{University of California, Berkeley}
  \city{Berkeley}
  \country{CA, USA}}
\email{yuxinc@berkeley.edu}

\author{Chen Tang}
\affiliation{
  \institution{The University of Texas at Austin}
  \city{Austin}
  \country{TX, USA}}
\email{chen.tang@austin.utexas.edu}

\author{Ran Tian}
\affiliation{
  \institution{University of California, Berkeley}
  \city{Berkeley}
  \country{CA, USA}}
\email{rantian@berkeley.edu}

\author{Chenran Li}
\affiliation{
  \institution{University of California, Berkeley}
  \city{Berkeley}
  \country{CA, USA}}
\email{chenran_li@berkeley.edu}

\author{Jinning Li}
\affiliation{
  \institution{University of California, Berkeley}
  \city{Berkeley}
  \country{CA, USA}}
\email{jinning_li@berkeley.edu}

\author{Masayoshi Tomizuka}
\affiliation{
  \institution{University of California, Berkeley}
  \city{Berkeley}
  \country{CA, USA}}
\email{tomizuka@berkeley.edu}

\author{Wei Zhan}
\affiliation{
  \institution{University of California, Berkeley}
  \city{Berkeley}
  \country{CA, USA}}
\email{wzhan@berkeley.edu}

\begin{abstract}

Generalization poses a significant challenge in Multi-agent Reinforcement Learning (MARL). The extent to which an agent is influenced by unseen co-players depends on the agent's policy and the specific scenario. A quantitative examination of this relationship sheds light on effectively training agents for diverse scenarios. In this study, we present the \emph{Level of Influence} (LoI), a metric quantifying the interaction intensity among agents within a given scenario and environment. We observe that, generally, a more diverse set of co-play agents during training enhances the generalization performance of the ego agent; however, this improvement varies across distinct scenarios and environments. LoI proves effective in predicting these improvement disparities within specific scenarios. Furthermore, we introduce a LoI-guided resource allocation method tailored to train a set of policies for diverse scenarios under a constrained budget. Our results demonstrate that strategic resource allocation based on LoI can achieve higher performance than uniform allocation under the same computation budget. 
\end{abstract}


\newcommand{\BibTeX}{\rm B\kern-.05em{\sc i\kern-.025em b}\kern-.08em\TeX}

\newcolumntype{L}[1]{>{\raggedright\let\newline\\\arraybackslash\hspace{0pt}}m{#1}}
\newcolumntype{C}[1]{>{\centering\let\newline\\\arraybackslash\hspace{0pt}}m{#1}}
\newcolumntype{R}[1]{>{\raggedleft\let\newline\\\arraybackslash\hspace{0pt}}m{#1}}

\setcopyright{none}
\renewcommand\footnotetextcopyrightpermission[1]{} 
\begin{document}


\pagestyle{fancy}
\fancyhead{}


\maketitle 


\section{Introduction}
\label{sec:Intro}
Creating agents capable of effectively interacting with other agents, in particular humans, has been a longstanding challenge~\cite{bard2020hanabi,dafoe2020open}. Agents trained with model-free reinforcement learning (RL) have shown the potential to reach or surpass human-level performance through self-play (SP) in both classical discrete board games~\cite{silver2017mastering,silver2018general,zha2021douzero} and continuous domains such as Dota~\cite{berner2019dota}, Starcraft~\cite{vinyals2019grandmaster}, and racing~\cite{fuchs2021super}. However, SP agents typically undergo training with replicas of themselves, resulting in limited adaptability and robustness when interacting with previously unseen co-players exhibiting different behaviors~\cite{lowe2020interaction,bullard2020exploring,strouse2021collaborating,mckee2022quantifying}.

One promising solution to improve the policy robustness is \emph{diversifying the co-player distribution}. It has been shown that computationally hard problems like chess and go could benefit from diversifying agents during training~\cite{zahavy2023diversifying}. Prior studies introduced and validated methods for more complex games, such as population-based training~\cite{jaderberg2019human, carroll2019utility,jaderberg2017population}, league-based training~\cite{vinyals2019grandmaster}, fictitious self-play~\cite{heinrich2015fictitious,strouse2021collaborating}, and diversification of agent hyperparameters~\cite{hu2020other,mckee2020social}. However, it is important to note that a trade-off exists in most of these methods, where enhancing generalization capabilities comes at the cost of increased training resources and time.


Nevertheless, an important question remains: \textit{is diversifying the co-player distribution during training always worthwhile?} In practice, various real-world applications require a set of tailored RL policies for diverse target  scenarios~\cite{lowe2017multi,fuchs2021super}. Diversifying the co-player distribution during training in all the target scenarios comes at a high training cost, with the resulting benefits varying based on the specific scenario. In particular, we argue that the benefits of introducing diverse co-players depend on how \emph{interactive} a scenario is.
For instance, consider training agents for autonomous driving tasks. Enhanced generalization does not provide substantial advantages on highways as it does in crowded intersections and roundabouts. On highways, vehicles primarily focus on maintaining a straight path and constant speed, involving fewer interactions. In contrast, in roundabouts and intersections where agents' paths are highly interdependent, the presence of diverse surrounding agent behaviors has a much more significant impact on the ego agent policy.

\textit{Our key insight is that, by quantifying the interaction intensity across various scenarios, we can assess the necessity of diversifying co-player policy distribution when training the ego agent policy as the effects of environmental variation, and allocate the training resources strategically to maximize the overall advantage.}

We introduce the \emph{Level of Influence} (LoI), a metric quantifying the interaction intensity among agents within a given scenario. We propose to quantify the interaction intensity by how much the ego agent's reward is affected by the variation of non-ego agents' behavior. Formally, inspired by~\cite{jaques2019social}, we define LoI as the mutual information (MI) between the ego agent's expected reward and the non-ego agent's policy selection. We validate the effectiveness of using LOI for cost-efficient generalization by training a set of policies with co-players of different diversities for groups of scenarios and environments. We find that the LoI metric is highly correlated with the benefits of having diverse co-player policy distribution on the generalization of the ego agent within given scenarios, i.e., a higher LoI value indicates a larger improvement can be achieved in the ego agent's performance when a more diverse set of co-play agents are encountered during training. 

Consequently, we design a LoI-guided resource allocation method to train a set of policies for diverse scenarios under a limited training budget. We compare the overall performance between the LoI-guided and uniform allocation schemes, showcasing that the LoI-guided scheme consistently yields higher average performance across a range of game settings.

We summarize the novel contributions of this paper as follows:
\begin{enumerate}[label=\arabic*.]
    \item We propose a novel metric, Level of Influence (LoI), to quantify the interaction among agents in general multi-agent reinforcement learning problems. 

    \item We demonstrate that the LoI metric is highly correlated with the benefits of having diverse co-player distribution on the generalization of the ego agent within given scenarios. 

    \item We propose a LoI-guided resource allocation method and show that it can achieve a higher average reward than uniform allocation under the same computation budget. 
\end{enumerate}


\section{Related Work}
\label{sec:RelatedWork}
\textbf{Ad-hoc Teamwork}. Ad-hoc teamwork (AHT)~\cite{stone2010ad}, also referred to as zero-shot coordination (ZSC)~\cite{hu2020other}, involves training agents to collaborate with co-players they have not encountered before. Early approaches primarily focused on game-theoretic analysis within matrix games~\cite{stone2009leading, agmon2012leading}. Recently, multi-agent reinforcement learning (MARL) has enabled ad-hoc teamwork in more intricate grid worlds and continuous domains. Various works have explored hierarchical social intention~\cite{kleiman2016coordinate}, social conventions~\cite{shih2021critical}, shared planning~\cite{ho2016feature}, and theory of mind~\cite{choudhury2019utility} in this context. In MARL, an agent's learning is influenced by both other co-players and the environment~\cite{littman1994markov}. However, most of the previously mentioned works do not explicitly evaluate the impacts of environmental variations. Carroll et al.~\cite{carroll2019utility} introduce the game Overcooked and explicitly showcase that the environment configurations affect the robustness of the trained agents when teamed with unknown human players. Subsequent research includes diverse layout generation~\cite{fontaine2021importance, mckee2022quantifying} and scalable evaluation~\cite{leibo2021scalable}. Nonetheless, these works only provide \emph{qualitative} analyses of different environments and do not \emph{quantitatively} measure such effects across scenarios.

\textbf{Generalization in Multi-agent RL.} In the field of MARL, various attempts have been made to enhance agents' adaptability to new co-players.
Jaderberg et al.~\cite{jaderberg2017population,jaderberg2019human} introduced population-based training (PBT) to jointly optimize the performance of a population of models. Several variations of PBT include the league-based training that masters the full game of StarCraft II~\cite{vinyals2019grandmaster}, the fictitious co-play (FCP) that can reach human-level performance~\cite{heinrich2015fictitious, strouse2021collaborating}, and heterogeneous populations training with Social Value Orientation (SVO)~\cite{mckee2020social}. However, high-performing agents come at the cost of more expensive training cost. Considering the varying benefits of generalization across diverse environments, we aim to evaluate if the extra training cost for enhanced generalization is justified. This area is relatively under-explored in existing research.

\textbf{Causal Influence.} Our work shows a notable connection to Jaques et al.~\cite{jaques2019social}, where a causal influence reward is incorporated as an intrinsic motivation during the training of RL agents. This reward incentivizes agents to maximize mutual information (MI) between their actions. The goal of maximizing MI between actions is to encourage more coordinated behavior among the agents. The causal influence is assessed using counterfactual reasoning~\cite{mcallister2022control,foerster2018counterfactual,pearl2013structural} where an agent simulates alternate, counterfactual \emph{actions} that it could have taken at every time step. In contrast, we measure the mutual information between the ego agent's expected reward and the non-ego agent's policy selection by simulating counterfactual \emph{policies} that the non-ego agent would have chosen within given scenarios.


\section{Preliminaries}
\label{sec:Preliminaries}

\subsection{Multi-agent MDP}
\label{sec:MDP}

An \emph{n-player partially observable Markov game} $\mathcal{M}$ \cite{boutilier1996planning, mckee2022quantifying} is defined by tuple $\mathcal{M} = \langle\mathcal{S},\mathcal{O},\{\mathcal{A}_i\}_{i\in\alpha},\mathcal{T},\{r_i\}_{i\in\alpha}\rangle$, where $\mathcal{S}$ is the finite set of \emph{states}, $\mathcal{O}\colon\mathcal{S}\times\{1,\dots,n\}\mapsto\mathbb{R}^d$ is the \emph{observation} function specifying each agent's $d$-dimensional view on the state space followed by their joint observation $\vec{o}=(o_1,\dots,o_n)$. Let $\alpha$ be a finite set of \emph{agents}, $\mathcal{A}_i$ is a finite set of discrete \emph{actions} available to agent $i$. The joint action is defined as $\vec{a}=(a_1,\dots,a_n)\in\mathcal{A}_1\times\cdots\times\mathcal{A}_n$. The stochastic \emph{transition function} $\mathcal{T}\colon\mathcal{S}\times\mathcal{A}_1\times\cdots\times\mathcal{A}_n \mapsto \Delta(\mathcal{S})$ determines the discrete probability distribution over the next state given the current state and the joint action. Each agent $i$ receives its real-valued \emph{reward} defined as $r_i\colon\mathcal{S}\times\mathcal{A}_1\times\cdots\times\mathcal{A}_n\mapsto\mathbb{R}$.

Each agent learns their policy in a \emph{decentralized} manner (\textit{i.e.}, independently learns a \emph{policy} $\pi_i(a_i|o_i)$ based on its own observation $o_i$ by optimizing its own individual reward $r_i$) without direct communication with other agents. We use $\vec{\pi}=(\pi_1,\dots,\pi_n)$ to denote the joint policy. Let $\gamma$ denote the discount factor that discounts future reward, agent $i$ optimizes for a policy that maximizes the long-term $\gamma$-discounted payoff \cite{mckee2022quantifying} defined as
\begin{equation}
    \label{eq:prelim_MDP}
    V_{\pi_i}(s_0) = \mathbb{E}\left[\sum_{t=0}^\infty \gamma^t r_i(s_t,\vec{a}_t)|\vec{a}_t\sim\vec{\pi}_t,s_{t+1}\sim\mathcal{T}(s_t,\vec{a}_t)\right].
\end{equation}

\subsection{Multi-agent Reinforcement Learning}
\label{sec:MARL}
\textbf{Self-play.} Self-play (SP) is an online evolutionary algorithm in which agents learn by playing against duplicates of themselves. Policies trained via SP have succeeded in a variety of environments and game settings \cite{silver2018general, vinyals2019grandmaster, berner2019dota, zha2021douzero}. In the SP training, all the agents are initialized with random policies and we keep updating the ego agent's policy while fixing other agents' policies. Throughout the training phase, the policies of the ego agent are stored as checkpoints periodically. Subsequently, following each checkpoint save, all non-ego agents adopt the recently saved checkpoint as their updated policies. (\textit{i.e.}, all non-ego agents become the latest duplicates of the ego agent). One major drawback of SP is that the learning agent can not generalize well to new partners outside of its own policy distribution \cite{strouse2021collaborating,bullard2020exploring,bullard2021quasi,lowe2020interaction} as agents only learn how to collaborate with themselves during training.

\textbf{Population-play.} Population-play (PP), on the other hand, keeps a population of agents training in parallel \cite{jaderberg2017population}. The environment and its agents are initialized with $p$ different random seeds ($p$ is the number of \emph{population}), and each population follows the same training scheme as SP. Instead of loading the recently saved checkpoints from their own population, non-ego agents of each population randomly select a population and adopt this population's latest checkpoint as their updated policies. By mutating different checkpoints across multiple populations, PP agents acquire better generalization capabilities than SP as a wider range of behavioral distribution has been covered during training \cite{jaderberg2017population, carroll2019utility, strouse2021collaborating, mckee2022quantifying}.

\textbf{Training Algorithm.} In multi-agent RL, agents' policies are parameterized by neural network models trained with various deep RL algorithms. In our case, we use Proximal Policy Optimization (PPO) \cite{schulman2017proximal} for model training.


\section{Methodology}
\label{sec:Methods}
We aim to study the potential impact of the diversity of co-play agents during training on the generalization performance as the effects of environmental variation. In MARL, the performance of a policy is measured by its expected reward when pairing with different co-players, which is determined by the reward design (\textit{i.e.}, payoff matrix). We use the term \emph{environments} to refer to games that have distinct reward designs. Under the same reward design, one can create different environmental variations by changing the map layout (\textit{e.g.}, map size, map shape, obstacle locations, and more). We use the term \emph{scenarios} to refer to distinct map layouts within the same \emph{environment}. Enhanced generalization yields varied levels of performance improvements for the agent in different scenarios. Intuitively, this is because agents in different scenarios have different interaction intensities, and the generalization performance as measured by the expected reward depends greatly on the interaction pattern and frequency. Therefore, we aim to find a metric that quantifies the interaction between agents as a way to predict the potential generalization improvement by having a more diverse set of co-player policies during training.

\begin{algorithm}
    \caption{Level of Influence calculation}
    \label{alg:LoI}
    \begin{algorithmic}[1]
        \Require \# Alice policies $a$, \# Bob policies $b$, \# Alice checkpoints per policy $m$, \# Bob checkpoints per policy $n$, Alice sampling probability $P_{\varphi}$, Bob sampling probability $P_{\vartheta}$, \# game per Alice-Bob pair $g$
        \State Train $a$ Alice policies with SP and save checkpoints pool $\Phi_i$
        \State Train $b$ Bob policies with SP and save checkpoints pool $\Theta_j$
        \State Initialize set of index $\mathcal{I} \gets \{\}$
        \For{i=1:a}
            \State Sample $m$ Alice checkpoints $\phi_{i,k}\sim \Phi_i$ with $P_\varphi$
            \For{k=1:m}
                \For{j=1:b}
                \State Sample $n$ Bob checkpoints $\theta_{j,l}\sim \Theta_{j}$
                with $P_\vartheta$
                \State Initialize set of distribution $\mathcal{P} \gets \{\}$
                    \For{l=1:n}
                        \State $P_{R_{i,k}|\vartheta=\theta_{j,l},~\varphi=\phi_{i,k}}\gets$ \Call{Distribution}{$\phi_{i,k}$,~$\theta_{j,l}$,~$g$}
                        \State $\mathcal{P} \gets \mathcal{P}\cup P_{R_{i,k}|\vartheta=\theta_{j,l},~\varphi=\phi_{i,k}}$
                    \EndFor
                    \State $P_{R_{i,k}|\varphi=\phi_{i,k}} \gets $\Call{Marginal\_Distribution}{$\mathcal{P}$,~$P_{\vartheta}$}
                    \Statex\Comment{Eq.~(\ref{eq:marg_PMF})}
                    \State $I_{i,j,k} \gets $\Call{Mutual\_Information}{$P_{R_{i,k}|\varphi=\phi_{i,k}}$,~$\mathcal{P}$}
                    \Statex\Comment{Eq.~(\ref{eq:LoI})}
                    \State $\mathcal{I} \gets \mathcal{I}\cup I_{i,j,k}$
                \EndFor
            \EndFor
        \EndFor
        \State $\bar{I} \gets$ \Call{Average}{$\mathcal{I}$}
        \Ensure $\bar{I}$
        
        \Function{Distribution}{$\phi$,~$\theta$,~$g$}
            \State Initialize set of reward $\mathcal{R} \gets \{\}$
            \For{i=1:g}
                \State Game between $\phi$ and $\theta$ to collect reward $r$
                \State $\mathcal{R} \gets \mathcal{R} \cup r$
            \EndFor
            \State $P_{R|\vartheta=\theta,\varphi=\phi} \gets$ \Call{Histogram}{$\mathcal{R}$}
            \State \Return $P_{R|\vartheta=\theta,\varphi=\phi}$
        \EndFunction
    \end{algorithmic}
\end{algorithm}

\subsection{Level of Influence}
\label{sec:LoI}
In order to quantitatively describe the interaction intensity between each agent in a certain \emph{scenario} as its intrinsic property, we take inspiration from causal influence~\cite{jaques2019social} and define a new metric named \emph{Level of Influence} (LoI). For simplicity, consider a symmetric game with two agents named Alice and Bob. Alice is the algorithm-controlled agent (\textit{i.e.}, ego agent), and Bob can be another algorithm-controlled agent with an unknown policy or human player (\textit{i.e.}, non-ego agent). We would like to quantify the expected impact of Bob's behavior on Alice's performance within this scenario.

Suppose Alice and Bob are algorithm-controlled agents with policy $\phi\in\Phi$ and $\theta\in\Theta$, respectively, where $\Phi$ is a policy class of size $m$ and $\Theta$ is a policy class of size $n$. Alice and Bob choose their policy following the policy distribution of $P_\varphi(\phi)=\mathbb{P}[\varphi=\phi]$ and $P_\vartheta(\theta)=\mathbb{P}[\vartheta=\theta]$. Let $r\in\mathbb{R}$ denote the total reward of Alice paired with Bob. Under the two-agent game setting, Alice's reward is affected by both agents' policy choices, and the conditional reward distribution of Alice given Alice's policy $\varphi=\phi$ and Bob's policy $\vartheta=\theta$ can be represented as
\begin{equation}
\label{eq:cond_PMF}
    P_{R|\vartheta,\varphi}(r|\theta,\phi) = \mathbb{P}[R=r|\vartheta=\theta,\varphi=\phi].
\end{equation}
Here we use $R$, $\varphi$ and $\vartheta$ to denote the random variable of $r$ $\phi$, and $\theta$, respectively. We can then get Alice's marginal reward distribution as
\begin{equation}
\label{eq:marg_PMF}
    P_{R|\varphi}(r|\phi) = \sum_{\theta\in\Theta} P_{R,\vartheta|\varphi}(r,\theta|\phi) = \sum_{\theta\in\Theta} P_{R|\vartheta,\varphi}(r|\theta,\phi)P_{\vartheta|\varphi}(\theta|\phi).
\end{equation}

The discrepancy between the marginal reward distribution and the conditional reward distribution is a measure of the interaction level. Intuitively, we want the LoI to measure the degree to which Alice's reward distribution changes induced by Bob's policy choice, given Alice's own policy choice. Therefore, the LoI is defined as the conditional mutual information of Alice's reward and Bob's policy with respect to Alice's policy:
\begin{subequations}
\begin{align}
    I(R;\vartheta|\varphi) &=     \mathbb{E}_\varphi\left[D_\mathrm{KL}\left(P_{R,\vartheta|\varphi}\|P_{R|\varphi}P_{\vartheta|\varphi}\right)\right]\\
    &= \sum_{\phi\in\Phi}P_\varphi(\phi)D_\mathrm{KL}\left(P_{R,\vartheta|\varphi=\phi}\|P_{R|\varphi=\phi}P_{\vartheta|\varphi=\phi}\right)\\
    &= \sum_{\phi\in\Phi}P_\varphi(\phi)\mathbb{E}_\vartheta\left[D_\mathrm{KL}\left(P_{R|\vartheta,\varphi=\phi}\|P_{R|\varphi=\phi}\right)\right]\\
    &= \sum_{\phi\in\Phi}P_\varphi(\phi)\sum_{\theta\in\Theta}P_\vartheta(\theta)D_\mathrm{KL}\left(P_{R|\vartheta=\theta,\varphi=\phi}\|P_{R|\varphi=\phi}\right).
\label{eq:LoI}
\end{align}
\end{subequations}
It is worth noting that when $I(R;\vartheta|\varphi)=0$, Alice's total reward will not be affected by Bob's policy choice at all under the given scenario, thus there is little value for training Alice's policy with diverse opponent's policy. The higher this value becomes, the more significant impact Bob's policy will have on Alice's expected reward; consequently, encountering a more diverse Bob's policy when training Alice's policy can help improve performance when paired with unseen partners and more training budget is well justified.
\begin{figure*}[htbp!]
    \centering
    \subcaptionbox{Small
    \label{fig:config_small}}[0.9in][c]{%
        \includegraphics[height=1.2in]{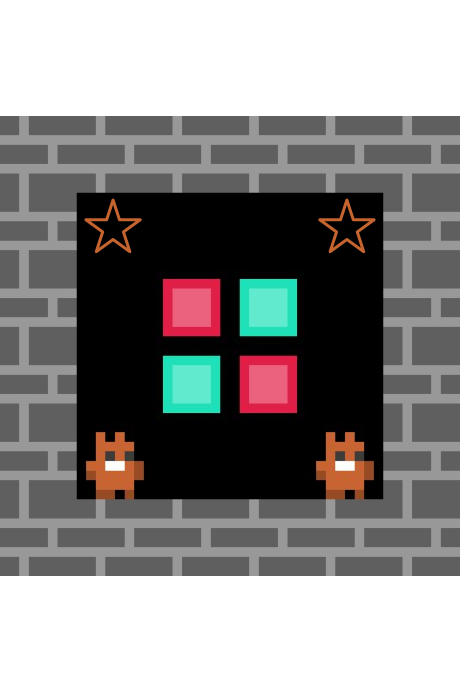}}
    \subcaptionbox{Medium
    \label{fig:config_medium}}[1.1in][c]{%
        \includegraphics[height=1.2in]{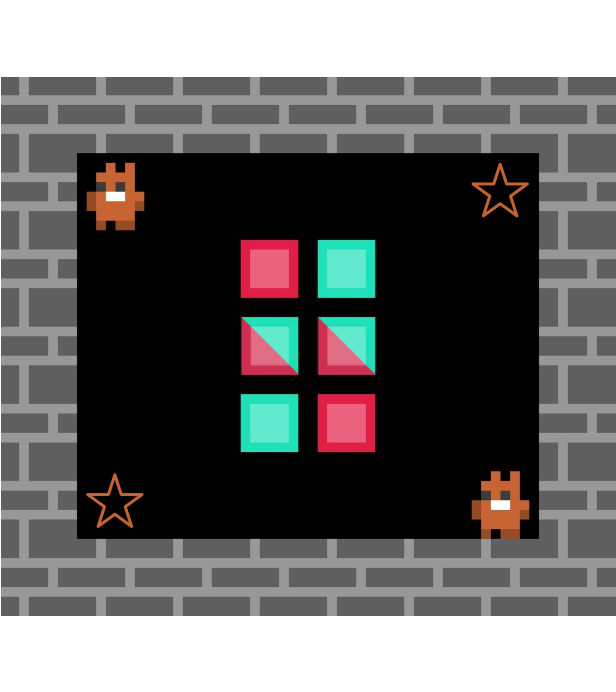}}
    \subcaptionbox{Large
    \label{fig:config_large}}[1.8in][c]{%
        \includegraphics[height=1.2in]{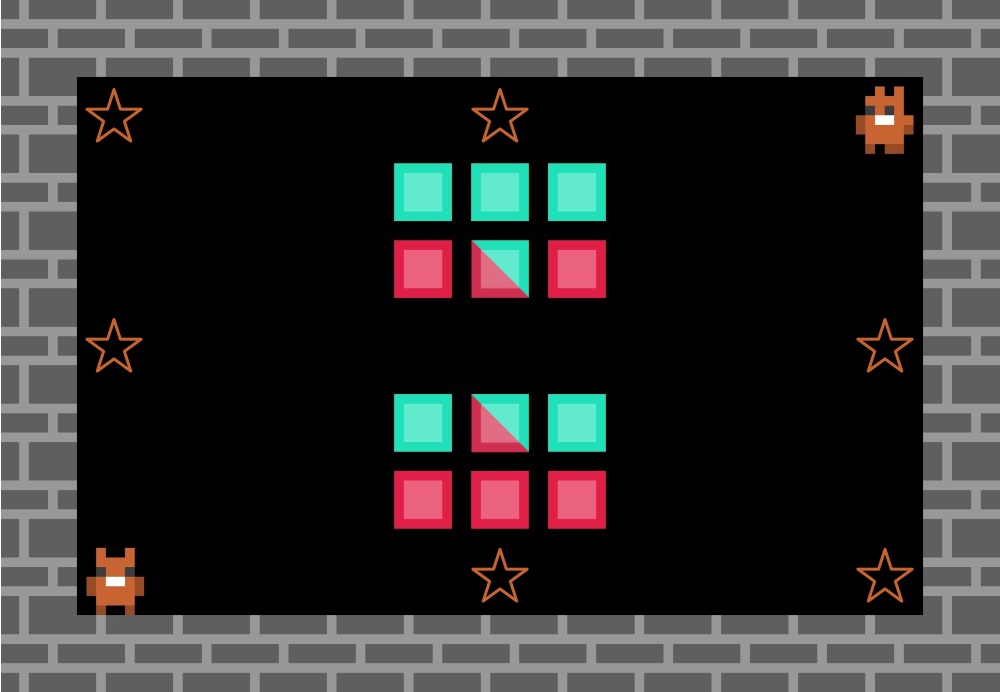}}
    \subcaptionbox{Obstacle
    \label{fig:config_obstacle}}[1.8in][c]{%
        \includegraphics[height=1.2in]{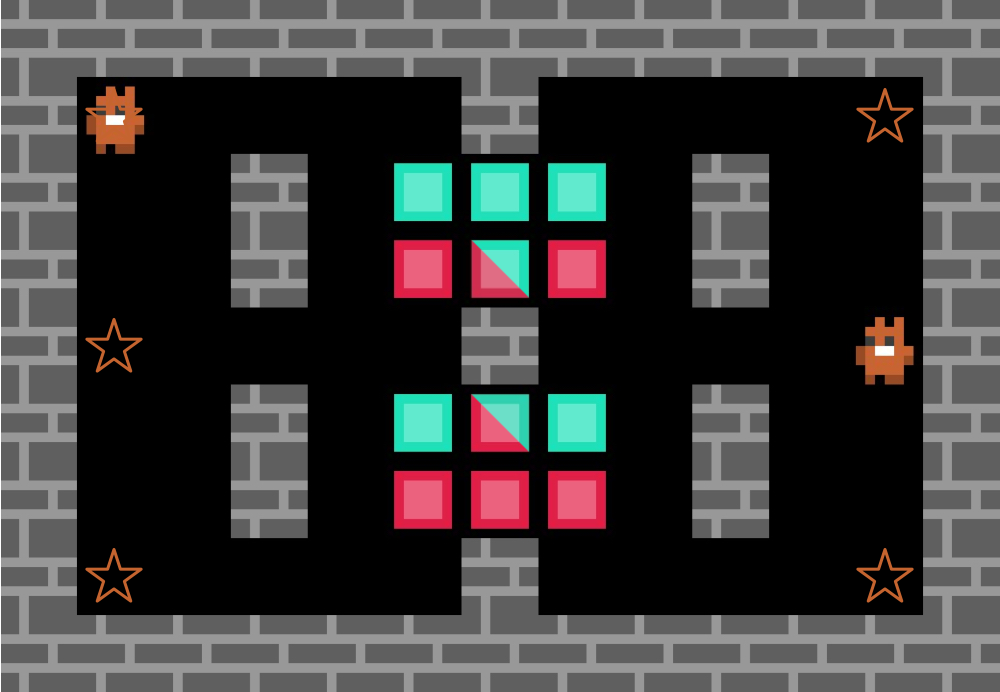}}
    \caption{We investigate the influence between agents under \textit{$^\star$ in the Matrix} game setting across four distinct 2-player scenarios: (a) Small, (b) Medium, (c) Large, and (d) Obstacle. Brown hollow stars denote the random spawn spots for both agents. Single-color blocks (cyan and red) denote the fixed resource, and mixed-color blocks denote the random resource, which can be changed during each initialization.}
    \label{fig:configs}
\end{figure*}

\subsection{Approximation}
\label{sec:Approx}
Since we do not have access to both agents' full policy space $\Phi$ and $\Theta$, it is computationally infeasible to calculate the ground truth LoI following Eq. (\ref{eq:LoI}). We need to approximate the LoI for the given scenario with a small amount of computational resources. In practice, we train $a+b$ SP policies, choose $a$ of them as Alice's policies, and the rest $b$ policies as Bob's policies, randomly. Training a convergent SP policy gives us a pool of checkpoints, including the early-, middle-, and late-stage generations, with various skill levels and collaborating patterns. We choose $m$ checkpoints from the late stage of each Alice policy as a group of Alice with slightly different skills and choose $n$ checkpoints from all stages of each Bob policy as samples of Bob's policy distribution. We summarize our LoI calculation in Algorithm~\ref{alg:LoI}.


\section{Environments}
\label{sec:Env}
We adopt DeepMind \emph{Melting Pot} environment \cite{MeltingPot} for evaluation. \emph{Melting Pot} is a multi-agent RL environment with different \emph{substrates} (\textit{i.e.}, physical environment) of zero-sum, shared-reward, and general-sum games. In our case, we choose the two-agent substrate named ``\emph{$^\star$ in the Matrix},'' whose mechanism is introduced in \cite{vezhnevets2019options}. In this game, two agents can move around the map, collect $K$ different resources, and fire ``interaction beams.'' Each agent has its inventory $\rho\in\mathbb{R}^K$ to track the number of resources of each kind picked up since the last re-spawn, i.e., $\rho_i$ denotes the number of the $i^{\mathrm{th}}$ type of resources in its inventory. Inventory is only visible to themselves. 

An \emph{interaction} occurs whenever one agent zaps the other agent with their interaction beam. The interaction is then resolved by a matrix game with the payoff matrix $A$ describing the reward corresponding to the pure strategies of the matrix game available to each agent. Each kind of resource maps one-to-one to each pure strategy. During the interaction, each agent executes a mixed strategy depending on the resources they picked up before the interaction. In particular, an agent with inventory $\rho$ plays the mixed strategy with weights $\nu=(\nu_1,\dots,\nu_K)$, where $\nu_i=\rho_i/(\sum_{j=1}^K\rho_j)$. Intuitively, the more resources of a certain kind are picked up by an agent, the more likely this agent executes the corresponding strategy of that kind of resource. Formally, during the interaction, a pure strategy is sampled from each player's mixed strategy distribution defined by $\nu$. We represent the sampled strategies of the row and column players as two one-hot vectors, denoted by $r_\mathrm{row},r_\mathrm{col}\in \mathbb{R}^K$, respectively. Afterward, the rewards that the row and column players obtain from the interaction, denoted by $r_\mathrm{row}$ and $r_\mathrm{col}$ respectively, are assigned via
\begin{subequations}
\begin{align}
\label{eq:reward}
    r_\mathrm{row} &= \nu_\mathrm{row}^\top A_\mathrm{row} \nu_\mathrm{col},\\
    r_\mathrm{col} &= \nu_\mathrm{row}^\top A_\mathrm{col} \nu_\mathrm{col}.
\end{align}
\end{subequations}
After the interaction, both agents will re-spawn after 5 steps. Each game lasts for 2000 steps.

\begin{table}[t]
  \caption{Payoff matrices $\mathbf{A_\text{row}}$ of the two-player \textit{$^\star$ in the Matrix} game. All four games are symmetric with $\mathbf{A_\text{row}={A_\text{col}}^\top}$.}
  \label{tab:payoff_matrices}
  \begin{tabular}{cccc}\toprule
    \multirow{2}{*}{\textit{Chicken}} & \textit{Pure} & \textit{Prisoners} & \textit{Stag}\\ 
    & \textit{Coordination} & \textit{Dilemma} & \textit{Hunt}\\ \midrule
    $\begin{pmatrix}3&2\\5&0\end{pmatrix}$ &
    $\begin{pmatrix}1&0\\0&1\end{pmatrix}$ &
    $\begin{pmatrix}3&0\\5&1\end{pmatrix}$ &
    $\begin{pmatrix}4&0\\2&2\end{pmatrix}$\\
    \bottomrule
  \end{tabular}
\end{table}

By changing the underlying payoff matrix, we are able to change the game property of the substrate. Therefore, we define four different \emph{environments} with various game properties, namely \emph{Chicken}, \emph{Pure Coordination}, \emph{Prisoners Dilemma}, and \emph{Stag Hunt}~\cite{MeltingPot} with payoff matrices defined in Table~\ref{tab:payoff_matrices}. All four environments are symmetric with 2 types of resources ($K=2$) and we have $\mathbf{A}_\text{row}={\textbf{A}_\text{col}}^\top$.

For each environment, we create four different \emph{scenarios} by changing the size of the map as well as the layout of the resource and objects inside (see Figure \ref{fig:configs}). From \emph{Small} to \emph{Obstacle}, the map sizes are $6\times6$, $7\times 8$, $9\times13$, and $9\times13$, respectively. We anticipate that different map sizes may lead to varying levels of interaction intensity among agents. The observation window of each agent is $5\times 5$ square centered at the agent itself, which means agents in \emph{Small} is able to observe all the resource at any spawn location.


\section{Experiments Design}
\label{sec:Exp}
We design a series of experiments to validate the effectiveness of using LOI for cost-efficient generalization and demonstrate a useful application of LoI in guiding resource allocation for cost-efficient policy training. In particular, we would like to examine the following hypotheses. 

\textit{Hypothesis 1.} LoI is strongly correlated to the benefits of having diverse co-player distribution on the generalization of the ego agent within given scenarios (Section~\ref{sec:Validation}).

\textit{Hypothesis 2.} Under the same computation budget, the set of ego agents trained with LoI-guided resource allocation can achieve higher average performance than uniform allocation  (Section~\ref{sec:ResouceAllo}).

\subsection{Validating the Level of Influence}
\label{sec:Validation}
As outlined in Section~\ref{sec:LoI}, our objective is to utilize LoI to predict the benefits of having diverse co-player distribution during training on the generalization. To validate this idea, we first evaluate the impact of different co-player diversities on generalization performance within different scenarios (Section \ref{sec:FixedBob}). We then calculate the LoI of those scenarios (Section \ref{sec:LoICalc}), and find the correlation between the performance improvement and the LoI conditioned on diversity (Section \ref{sec:Corr}).

\subsubsection{Fixed-Bobs Evaluation.}
\label{sec:FixedBob}
First, we evaluate the impact of different co-player diversities on generalization performance within different scenarios. we train a set of policies with different co-player diversities that we want to compare across all environment-scenario combinations. We then assess these trained policies against a set of predetermined agents for evaluation. Specifically, we train one SP policy for 5M steps and save a new checkpoint every 200K steps. We select four checkpoints at 1.4M, 2.6M, 3.8M, and 5M steps as a fixed group of policies for evaluation, which we refer to as "\emph{Fixed-Bobs}".

Afterward, we train 5 instances of SP policies with different random seeds (SP), one set of PP policies with 3 populations (PP3), and one set of PP policies with 5 populations (PP5). All policies are trained for 10M steps. We choose the final checkpoint of each SP and each population of PP (\textit{i.e.}, checkpoints at 10M steps) and pair them with the aforementioned Fixed-Bobs. Each game lasts for 2000 steps and repeats 10 times. We evaluate each training method and report the average individual reward for each policy by aggregating results across all 10 games, four Fixed-Bobs policies, and all populations (or all seeds in the case of SP). To better compare the performance gap between training methods across different scenarios, we normalize the previous results by dividing each element by the corresponding reward value from SP of the same environment and scenario.

\begin{figure*}[t]
    \centering
    \subcaptionbox{Chicken
    \label{fig:meltingpot_heatmap_c}}[0.23\linewidth][c]{%
        \includegraphics[width=0.95\linewidth]{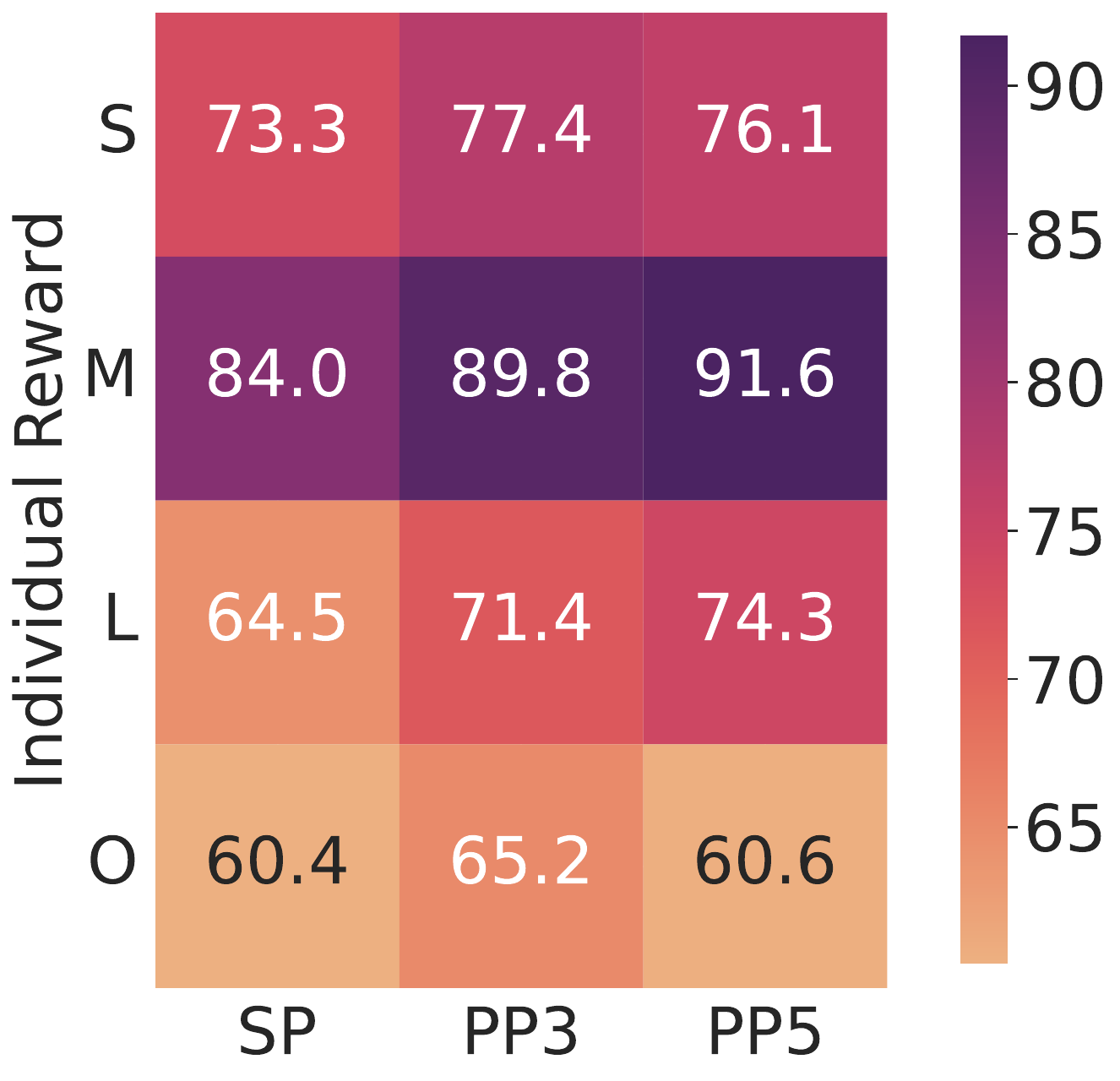}
        
        \includegraphics[width=0.95\linewidth]{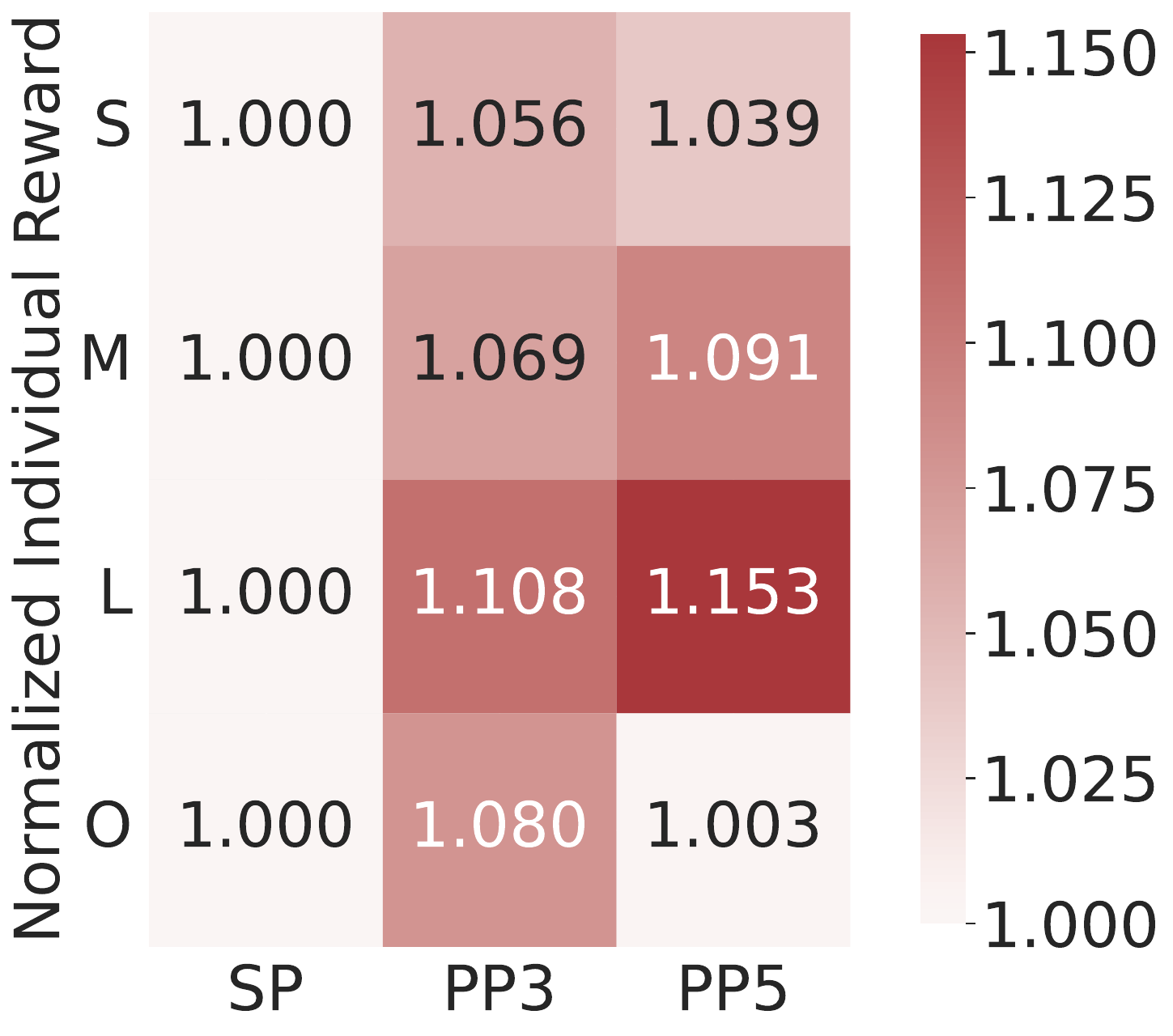}}\quad
    \subcaptionbox{Pure Coordination
    \label{fig:meltingpot_heatmap_pc}}[0.23\linewidth][c]{%
        \includegraphics[width=0.95\linewidth]{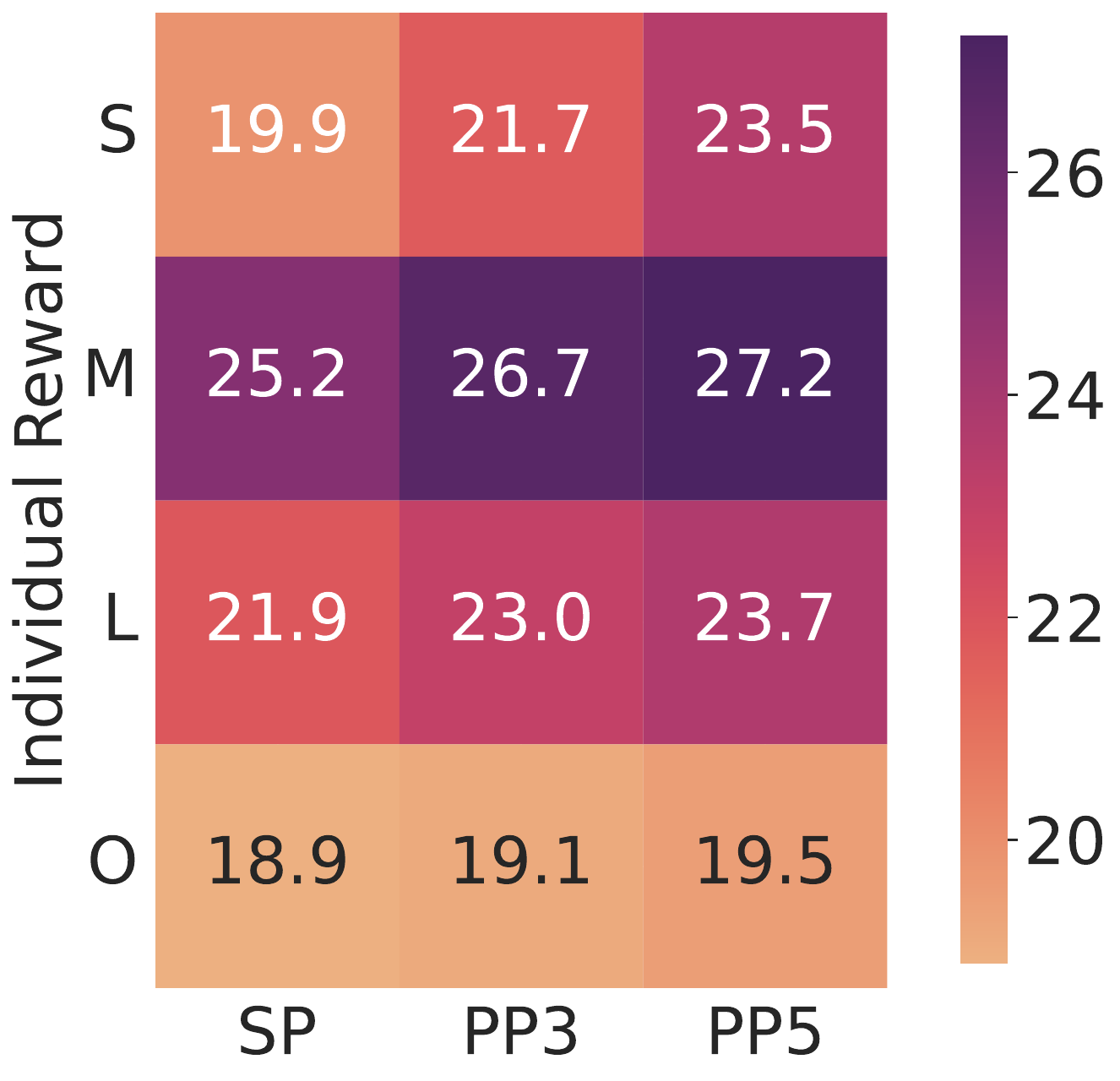}
        
        \includegraphics[width=0.95\linewidth]{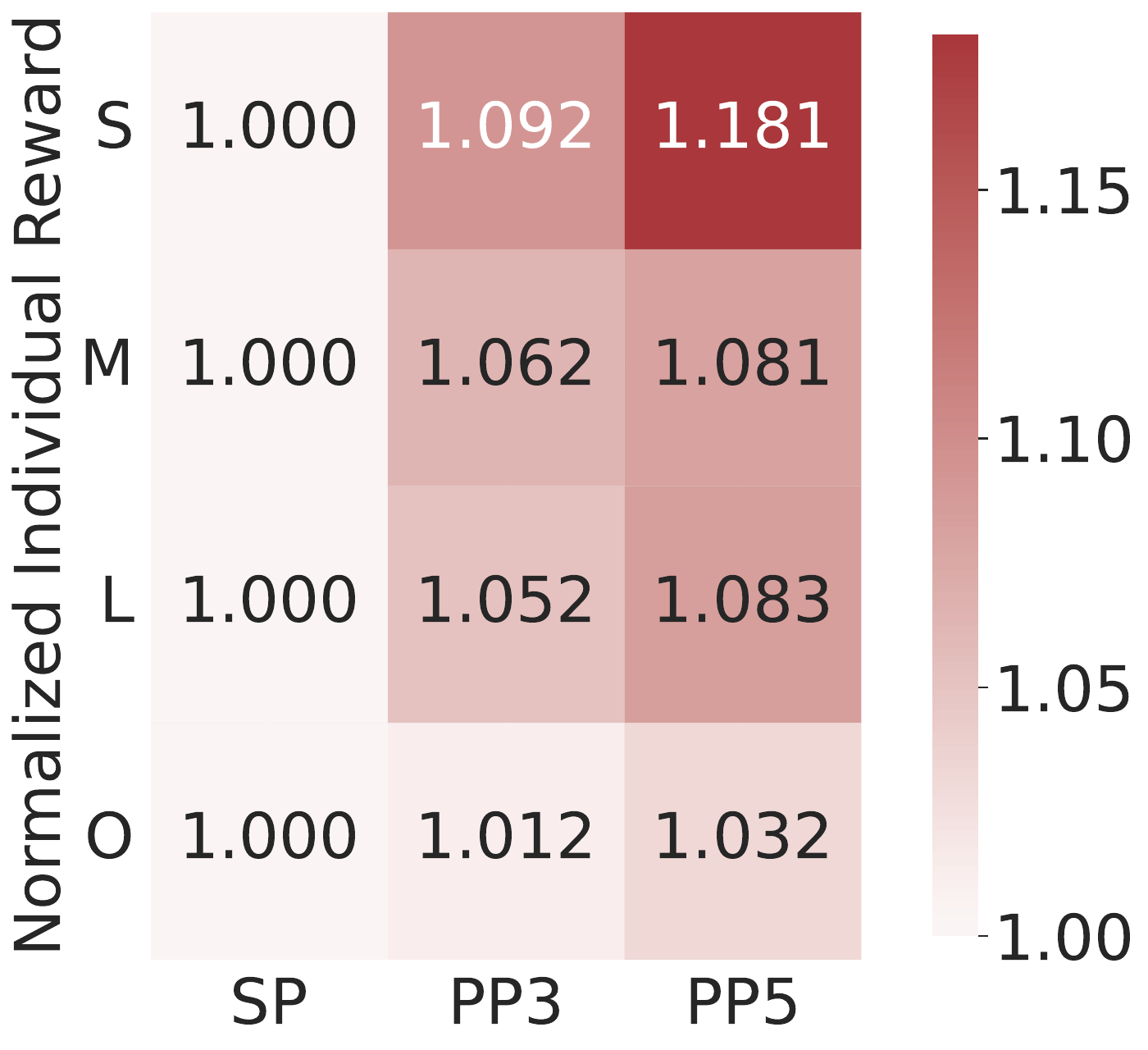}}\quad
    \subcaptionbox{Prisoners Dilemma
    \label{fig:meltingpot_heatmap_pd}}[0.23\linewidth][c]{%
        \includegraphics[width=0.95\linewidth]{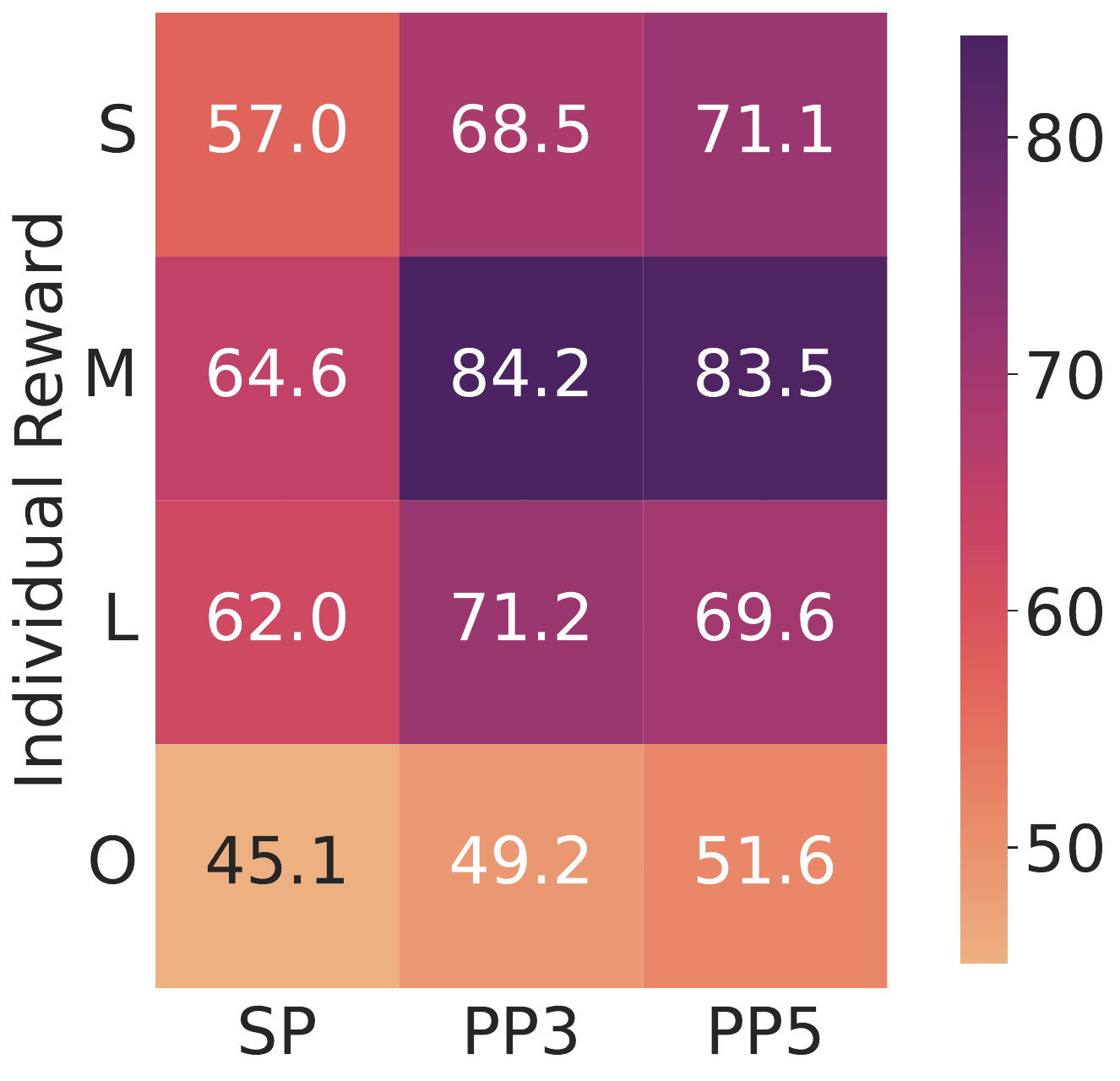}
        
        \includegraphics[width=0.95\linewidth]{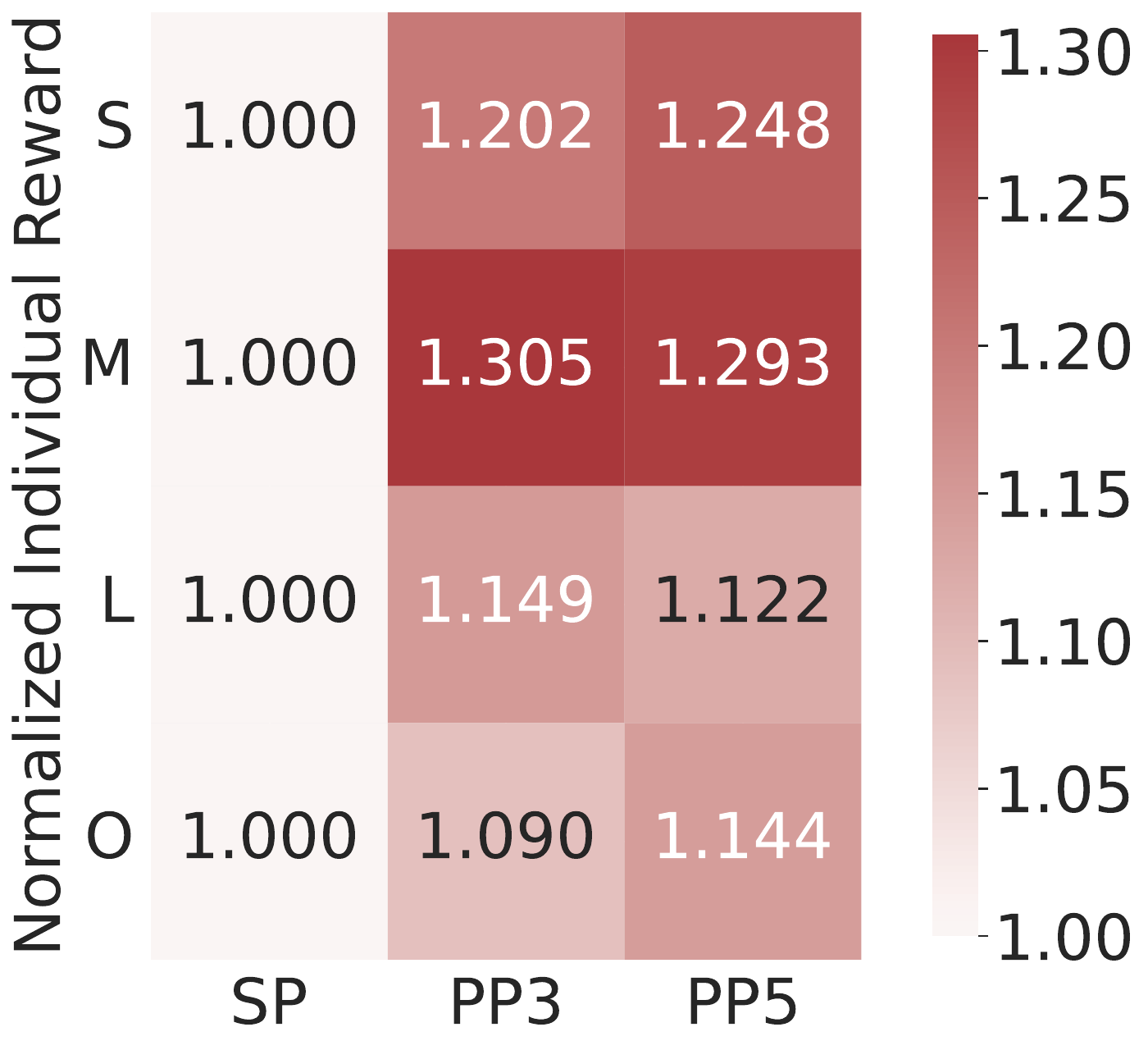}}\quad
    \subcaptionbox{Stag Hunt
    \label{fig:meltingpot_heatmap_sh}}[0.23\linewidth][c]{%
        \includegraphics[width=0.95\linewidth]{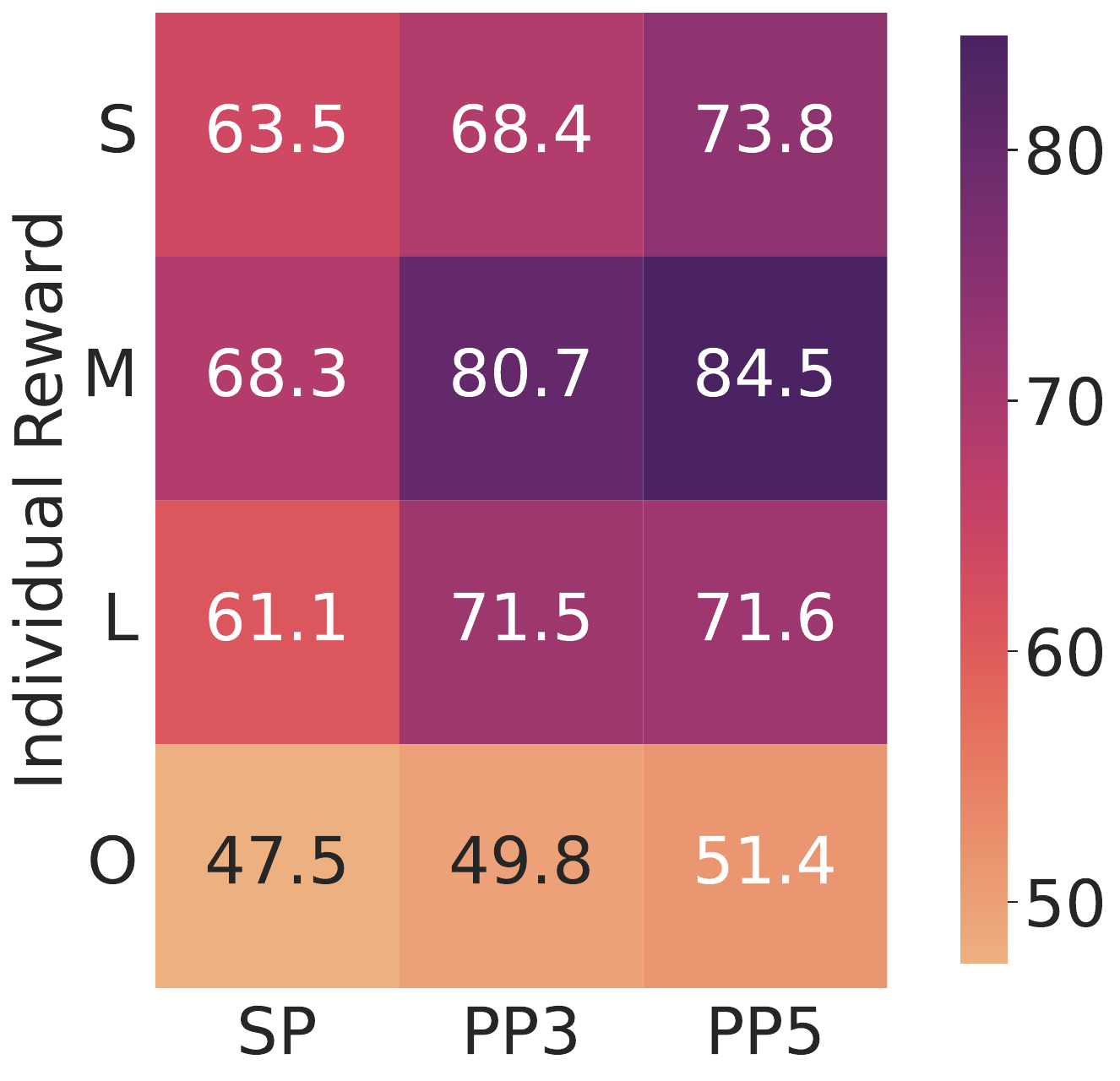}
        
        \includegraphics[width=0.95\linewidth]{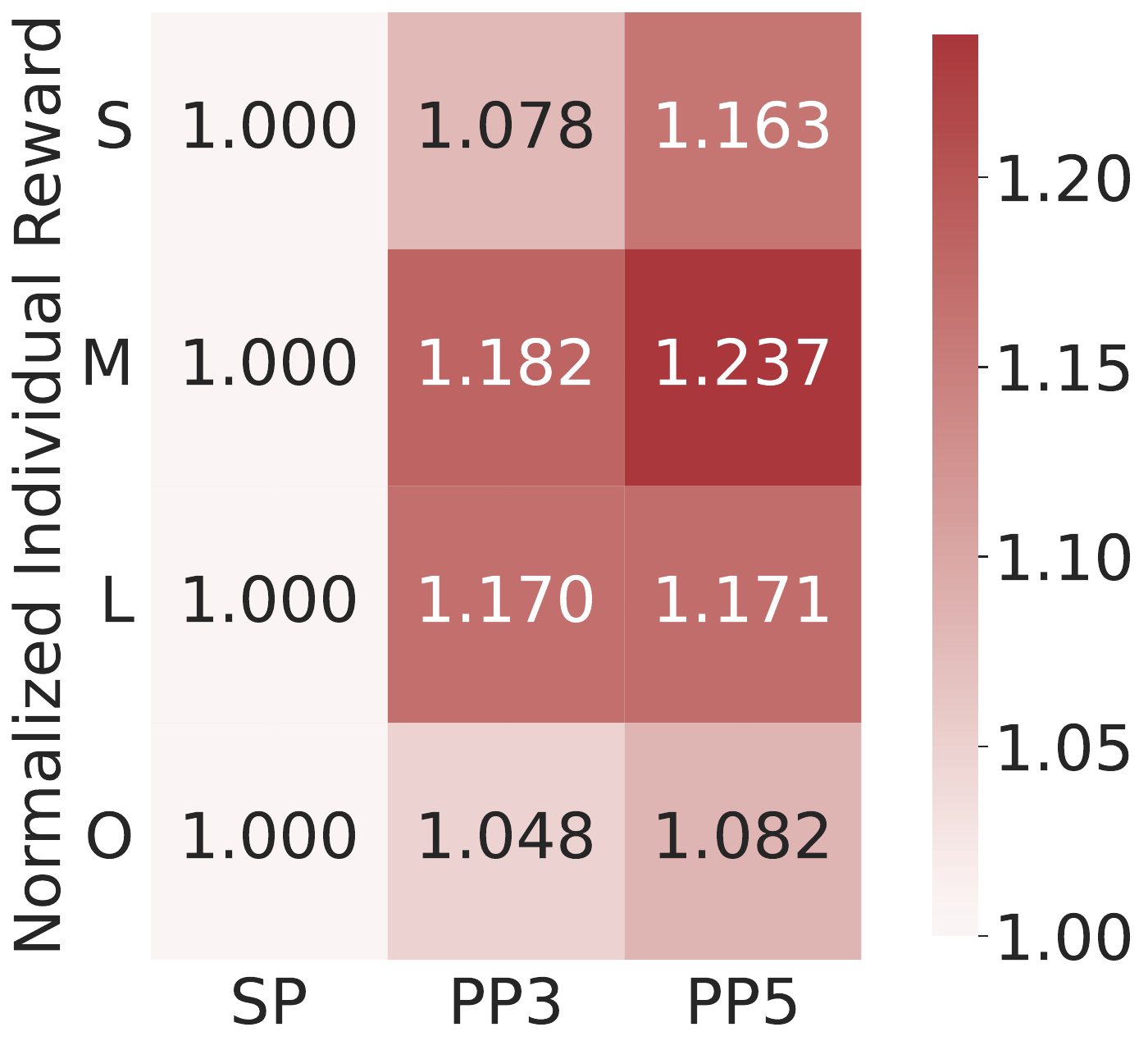}}
    \caption{\emph{Top row:} Fixed-Bobs evaluation of agents trained by self-play (SP), population-play $\mathbf{p=3}$ (PP3), population-play $\mathbf{p=5}$ (PP5) across four scenarios: Small (S), Medium (M), Large (L), and Obstacle (O). Each tile is the average over all populations and seeds (5 seeds for SP) with 10 independent games between each Alice-Bob combination. \emph{Bottom row:} Percentage of improvement calculated by dividing each row by its first element (SP). \emph{Result:} With a growing population, PP gains larger advantages over SP in general. However, the percentage of increase varies across different scenarios under the same environment, and the overall trend of improvement varies across different environments.}
    \label{fig:meltingpot_heatmap}
\end{figure*}

\subsubsection{LoI Calculation.}
\label{sec:LoICalc}
Second, we calculate the LoI introduced in Section~\ref{sec:Approx} and Algorithm~\ref{alg:LoI} for each scenario and environment. Specifically, we train 1 Alice policy ($a=1$) and 5 Bob policies ($b=5$) with different random seeds. We choose 4 late-stage generations at 3.8M, 4.2M, 4.6M, and 5M steps from Alice's checkpoint pool ($m=4$) and 9 all-stage generations at 0.2M, 0.6M, 1M, 1.4M, 1.8M, 2.6M, 3.4M, 4.2M, and 5M steps from Bob's checkpoints pools ($n=9$) of every policy. To compute the LoI, we model the policy distribution of Alice and Bob as a uniform distribution defined over their sampled checkpoints (\textit{i.e.}, $P_\varphi=1/4$ and $P_{\vartheta}=1/9$). We perform 6 games per Alice-Bob pair ($g=6$).

\subsubsection{Correlation between LoI and Average Improvement.}
\label{sec:Corr}
We also calculate the average improvement between each training method for each environment and scenario. Suppose the average individual reward for SP, PP3, and PP5 are $r_1$, $r_2$, and $r_3$, respectively, then the average improvement $\delta$ will be calculated as
\begin{equation}
\label{eq:avg_imp}
    \delta = \frac{r_2-r_1}{2} + \frac{r_3-r_2}{2} = \frac{r_3-r_1}{2}.
\end{equation}

Last, we find the correlation between the aforementioned LoI and average improvement on four scenarios within each environment. We apply the Pearson correlation coefficient as the measurement of linear correlation. Suppose the LoIs of four scenarios on the given environment are $I_i$ and the corresponding average improvements are $\delta_i$. Let $\bar{I}$ and $\bar{\delta}$ denote the mean LoI and mean average improvement over four scenarios, respectively. The correlation coefficient is then calculated as
\begin{equation}
\label{eq:corr_coeff}
    \gamma = \frac{\sum_{i=1}^4 (I_i-\bar{I})(\delta_i-\bar{\delta})}{\sqrt{\sum_{i=1}^4(I_i-\bar{I})^2}\sqrt{\sum_{i=1}^4(\delta_i-\bar{\delta})^2}}.
\end{equation}

\begin{table*}[t]
  \caption{One-way ANOVA tests on fixed-Bobs evaluation. Each ANOVA is summarized with an \textit{F}-statistic and a \textit{p}-value. \emph{Result:} Reported ANOVAs confirm that SP, PP3, and PP5 have significant effect on individual rewards in all four specified scenarios under four different environments.}
  \label{tab:pvalue}
  \begin{tabular}{ccccccccc} \toprule & \multicolumn{2}{c}{\textit{Chicken}} & \multicolumn{2}{c}{\textit{Pure Coordination}} & \multicolumn{2}{c}{\textit{Prisoners Dilemma}} & \multicolumn{2}{c}{\textit{Stag Hunt}} \\
  \cmidrule(lr){2-3} \cmidrule(lr){4-5} \cmidrule(lr){6-7} \cmidrule(lr){8-9} Scenario & \textit{F}-statistic & \textit{p}-value & \textit{F}-statistic & \textit{p}-value & \textit{F}-statistic & \textit{p}-value & \textit{F}-statistic & \textit{p}-value\\ \midrule 
    \textit{Small} & 8.338 & $3.46\times 10^{-4}$ & 238.9 & $5.19\times 10^{-51}$ & 72.48 & $1.01\times 10^{-23}$ & 97.51 & $2.83\times 10^{-29}$\\ 
    \textit{Medium} & 29.03 & $1.25\times 10^{-11}$ & 129.0 & $2.77\times 10^{-35}$ & 288.7 & $1.90\times 10^{-56}$ & 201.0 & $2.83\times 10^{-46}$\\ 
    \textit{Large} & 49.55 &  $8.16\times 10^{-18}$ & 67.61 & $1.53\times 10^{-22}$ & 75.46 & $1.99\times 10^{-24}$ & 132.2 & $7.40\times 10^{-36}$\\ 
    \textit{Obstacle} & 6.115 &  $2.70\times 10^{-3}$ & 5.495 & $4.84\times 10^{-3}$ & 43.89 & $3.31\times 10^{-16}$ & 20.80 & $7.72\times 10^{-9}$\\ 
    \bottomrule \end{tabular}
\end{table*}

\subsection{Resource Allocation}
\label{sec:ResouceAllo}
We can utilize the proposed LoI for allocating training resource allocation under a limited computation budget. Specifically, we would like to train a set of policies tailored to manage diverse scenarios within a consistent game setting, with the total computational resource fixed. Without extra information, one may distribute resources uniformly across all scenarios. If we attain LoI that bridges the expected performance improvement with additional training cost, we can allocate the resource accordingly, \textit{i.e.}, train policies with larger populations for scenarios with higher LoI and vice versa.

To demonstrate the proposed allocation strategy, we set a fixed training budget of 120M steps for training four policies, each handling a specific scenario. In the uniform allocation scenario, we train a 3-population PP policy (PP3) for 10M steps per seed on each scenario, summing up to 120M steps for all four scenarios. In the resource allocation approach, we calculate the LoI for each scenario (as outlined in Section~\ref{sec:LoICalc}) and devise a \emph{heuristic method} to allocate resources based on this metric. By default, we allocate 30M steps for each scenario (equivalent to the cost of training PP3 for 10M steps) and compute the mean LoI across the four scenarios. Scenarios with LoI greater than one standard deviation from the mean receive 50M steps (cost of training PP5 for 10M), while those with LoI less than one standard deviation get 10M steps (cost of training SP for 10M). Adjustments are made for scenarios with LoI within one standard deviation to maintain the total budget of 120M steps. For instance, if one scenario uses only 10M steps, the saved 20M steps are reallocated to the scenario with the highest LoI among the remaining three, and vice versa (see Appendix~\ref{app:methods} for further details).

We apply the Fixed-Bobs evaluation and compare the average normalized individual reward (see Section~\ref{sec:FixedBob}) over all four scenarios between uniform allocation and heuristic allocation.

\newpage
\section{Results}
In this section, we report the results and ablation studies of the experiments introduced in Section~\ref{sec:Exp}.
\subsection{Validating the Level of Influence}
\label{sec:ValidationResult}

\subsubsection{Fixed-Bobs Evaluation.}
\label{sec:FixedBob_result}

The full results of the Fixed-Bobs evaluation are shown in Figure~\ref{fig:meltingpot_heatmap}. Across all environments, individual rewards demonstrate an upward trend across scenarios as the training population size increases. Notably, we can regard SP as a specialized PP with a population size of 1. The absolute reward values differ significantly among diverse environments, strongly influenced by the unique game properties of each environment and the specific payoff matrix (Figure~\ref{fig:meltingpot_heatmap}, top row). This suggests that our scenario design gives rise to varying levels of interaction intensity among agents in all environments.

If we normalize the individual rewards according to Section~\ref{sec:FixedBob}, we get the percentage improvement to compare the generation performance between training methods (Figure~\ref{fig:meltingpot_heatmap}, bottom row). We can observe that the percentage improvement with increasing co-player diversity during training differs for each scenario within a specific environment, and the overall improvement trend varies across diverse environments.

In \emph{Chicken}, we note that an increase on population size provides the most substantial benefit in the Large scenario. However, this advantage is not as prominent in the Small scenario and diminishes significantly in the Obstacle scenario. In contrast, larger population size brings the highest improvement in the Small scenario of \emph{Pure Coordination} and progressively decreases in larger scenarios. \emph{Prisoners Dilemma} and \emph{Stag Hunt} share a similar trend where the benefits are particularly prominent in the Medium scenario, compared to the Obstacle one.

We perform the Analysis of Variance (ANOVA) analysis \cite{edwards2005ra} on the results. The ANOVA method examines whether there are significant differences in means among two or more groups. We report the $F$-statistic and a corresponding $p$-value with a null hypothesis that there is no noteworthy difference (Table~\ref{tab:pvalue}). We confirm that changing the population size (\textit{i.e.}, diversity of co-play agent's policy distribution) has a statistically significant effect on the generalization performance within different scenarios across all four environments. The results also validate that the significance of the effects varies across different scenarios for a given environment. 

\begin{table}[t]
\setlength{\tabcolsep}{2pt}
  \caption{LoI (and standard deviations, reported in parentheses) across four scenarios under four environments. Values are calculated over one Alice set ($\mathbf{m=4}$) and five Bob sets ($\mathbf{n=9}$) of different seeds, 10 independent games between each Alice-Bob combination. \emph{Result:} LoI exhibits varying trends across four specified scenarios in different environments.}
  \label{tab:index}
  \begin{tabular}{ccccc}\toprule
    & \multirow{2}{*}{\textit{Chicken}} & \textit{Pure} & \textit{Prisoners} & \textit{Stag}\\ 
    & & \textit{Coordination} & \textit{Dilemma} & \textit{Hunt}\\ \midrule
    \textit{Small} & 1.291 (0.14) & \textbf{1.117} (0.12) & 1.377 (0.11) & 1.397 (0.14)\\
    \textit{Medium} & 1.364 (0.09) & 1.071 (0.15) & \textbf{1.385} (0.11) & \textbf{1.431} (0.13)\\
    \textit{Large} & \textbf{1.438} (0.09) & 0.976 (0.09) & 1.180 (0.09) & 1.424 (0.07)\\
    \textit{Obstacle} & 1.227 (0.17) & 0.976 (0.18) & 1.100 (0.12) & 1.063 (0.11)\\
    \bottomrule
  \end{tabular}
\end{table}

\subsubsection{LoI Calculation.}

We calculate the mean LoI and standard deviation for each scenario and environment, as detailed in Section~\ref{sec:LoICalc}. The comprehensive results are presented in Table~\ref{tab:index}, with the maximum value in each environment highlighted in bold. It exhibits varying trends across four scenarios in different environments. In the \emph{Chicken} game, the four scenarios exhibit varying LoIs, with the highest observed in the Large scenario. In the \emph{Pure Coordination} game, the Large and Obstacle scenarios demonstrate similar LoIs, while the Small scenario has notably the highest LoI. For the \emph{Prisoner's Dilemma} game, the Small and Medium scenarios show comparable, significantly higher LoI values compared to the Large and Obstacle scenarios. In the \emph{Stag Hunt} game, the Small, Medium, and Large scenarios share similar LoI values, with the highest in the Medium scenarios, while the Obstacle scenario has a remarkably low LoI in comparison.

\begin{table}[t]
  \caption{Average improvement on individual reward between SP, PP3, and PP5 under each scenario and environment. \emph{Result:} The advantage of PP over SP varies across different scenarios, and the correlations between scenario and reward increment vary across different environments.}
  \label{tab:avg_improve}
  \begin{tabular}{ccccc}\toprule
    & \multirow{2}{*}{\textit{Chicken}} & \textit{Pure} & \textit{Prisoners} & \textit{Stag}\\ 
    & & \textit{Coordination} & \textit{Dilemma} & \textit{Hunt}\\ \midrule
    \textit{Small} & 1.4130 & \textbf{1.7986} & 7.0535 & 5.1652\\
    \textit{Medium} & 3.8312 & 1.0248 & \textbf{9.4688} & \textbf{8.0993}\\
    \textit{Large} & \textbf{4.9293} & 0.9117 & 3.7931 & 5.2341\\
    \textit{Obstacle} & 0.0789 & 0.3020 & 3.2389 & 1.9517\\
    \bottomrule
  \end{tabular}
\end{table}

\subsubsection{Correlation between LoI and Average Improvement.}
Based on the individual rewards in Figure~\ref{fig:meltingpot_heatmap}, we calculate the average improvement following Section~\ref{sec:Corr}. The average improvements obtained are presented in Table~\ref{tab:avg_improve}, with the highest value in each environment emphasized in bold. It is evident that the trends in each environment align closely with the LoI outcomes depicted in Table~\ref{tab:index}. This observation naturally sets the stage for the correlation analysis discussed in the subsequent section.

\begin{table}[t]
  \caption{Pearson correlation coefficient between average improvement of PP over SP and LoI. \emph{Result:} It exhibits a strong correlation between the average improvement of PP over SP and LoI under all four environments.}
  \label{tab:cor_coeff}
  \begin{tabular}{ccccc}\toprule
    & \multirow{2}{*}{\textit{Chicken}} & \textit{Pure} & \textit{Prisoners} & \textit{Stag}\\ 
    & & \textit{Coordination} & \textit{Dilemma} & \textit{Hunt}\\ \midrule
    Statistic & 0.98966 & 0.86309 & 0.93888 & 0.86631\\
    \bottomrule
  \end{tabular}
\end{table}

We calculate the correlation coefficient as in Section~\ref{sec:Corr}, and the results are shown in Table~\ref{tab:cor_coeff}. The correlation coefficient ranges from $-1$ to $1$. An absolute value of $1$ indicates a perfect linear relationship between two groups, with all data points falling on a line. The results highlight a strong positive correlation between the average improvement of PP over SP (increasing population size) and LoI across all four environments. Consequently, we can utilize LoI as a reference to predict whether implementing a more resource-intensive training method (e.g., PP with a large population size) will yield a substantial improvement in generalization over a more cost-effective training method (e.g., SP or PP with a small population size) in a given scenario. Note that this correlation is valid within a specific environment. Comparing two scenarios across different environments is not meaningful in this context, since the scale of the reward varies across environments. 


\begin{table}[t]
  \caption{Allocated training resource according to the heuristic method introduced in Section~\ref{sec:ResouceAllo}. Each column adds up to 120M steps.}
  \label{tab:res_alloc}
  \begin{tabular}{ccccc}\toprule
    & \multirow{2}{*}{\textit{Chicken}} & \textit{Pure} & \textit{Prisoners} & \textit{Stag}\\ 
    & & \textit{Coordination} & \textit{Dilemma} & \textit{Hunt}\\ \midrule
    \textit{Small} & 30M & 50M & 30M & 30M\\
    \textit{Medium} & 30M & 30M & 50M & 50M\\
    \textit{Large} & 50M & 30M & 30M & 30M\\
    \textit{Obstacle} & 10M & 10M & 10M & 10M\\
    \bottomrule
  \end{tabular}
\end{table}

\subsection{Resource Allocation}
\label{sec:ResouceAlloResult}
We utilize the LoI values provided in Table~\ref{tab:index} for the heuristic method described in Section~\ref{sec:ResouceAllo}. The allocated training resources for each scenario are outlined in Table~\ref{tab:res_alloc}. Specifically, 10M, 30M, and 50M training steps correspond to SP, PP3, and PP5, respectively. The total steps for each environment sum up to 120M, adhering to the training budget cap.

\begin{figure}[t]
    \centering
    \includegraphics[width=\linewidth]{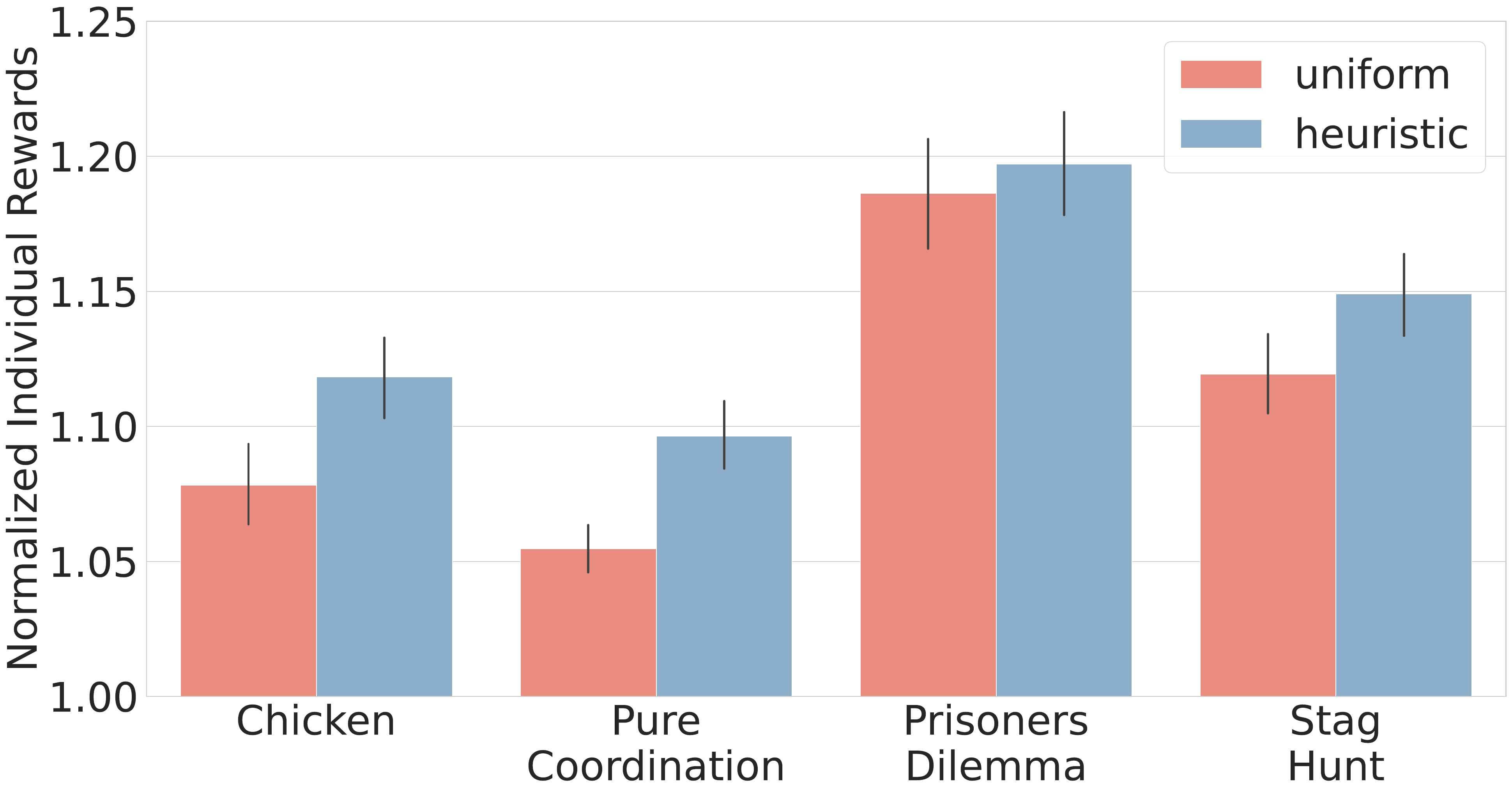}
    \caption{Fixed-Bobs evaluation comparison of the set of agents trained with uniformly allocated resource and LoI-guided heuristic allocation. Error bars correspond to $\mathbf{95\%}$ confidence intervals calculated over all populations across all four scenarios with 10 independent games between each Alice-Bob combination. \emph{Result:} The set of agents trained with LoI-guided resource allocation achieve higher overall performance under the same total resource budget cap within all four different environments.}
    \label{fig:resource_allocation}
\end{figure}

The comparison of Fixed-Bobs evaluation between uniform allocation and LoI-guided heuristic allocation is depicted in Figure~\ref{fig:resource_allocation}. Notably, the heuristic allocation demonstrates a substantial improvement in the average performance across all scenarios in the Chicken, Pure Coordination, and Stag Hunt environments.

We apply the two-sample one-tailed $t$-test \cite{student1908probable} for statistical analysis. It compares the means of two independent groups and determines if one is significantly larger than the other. We provide the $t$-statistic and a corresponding $p$-value with a null hypothesis that there is no noteworthy difference (Table~\ref{tab:resource_allocation}). We affirm that LoI-guided heuristic allocation exhibits a significant advantage over uniform allocation in all scenarios except for the Prisoners Dilemma, given the same resource budget cap. In conclusion, leveraging LoI enables us to strategically allocate resources for training a range of policies designed to handle diverse scenarios, resulting in improved overall performance within the same resource limit.

\begin{table}[t]
  \caption{Two-Sample One-tailed \textit{t}-Tests on resource allocation comparison. Each test is summarized with an \textit{t}-statistic and a \textit{p}-value. \emph{Result:} Reported tests suggest that except for the Prisoners Dilemma, LoI-guided heuristic allocation have a significant advantage over uniformly allocated case under the same total resource budget cap.}
  \label{tab:resource_allocation}
  \begin{tabular}{ccc}\toprule
    Scenario & \textit{t}-statistic & \textit{p}-value\\ \midrule
    \textit{Chicken} & 3.7649 & $1.050\times10^{-4}$\\
    \textit{Pure Coordination} & 5.3175 & $1.211\times10^{-7}$\\
    \textit{Prisoners Dilemma} & 0.7624 & $2.233\times10^{-1}$\\
    \textit{Stag Hunt} & 2.6498 & $4.297\times10^{-3}$\\
    \bottomrule
  \end{tabular}
\end{table}

It's important to highlight that the earlier discussed heuristic allocation is based on calculating LoI using checkpoints from 1 Alice policy and 5 Bob policies, with 5M steps per policy. Consequently, employing this heuristic resource allocation necessitates an additional 30M steps beyond the 120M steps budget ($25\%$ of the total budget). In Appendix~\ref{app:resource_alloc}, we show that, while augmenting the number of Bobs used in LoI computation significantly reduces the estimation variance, the proposed heuristic resource allocation scheme is less sensitive to the estimation noise. We can achieve comparable results as shown in Figure~\ref{fig:resource_allocation} with LoI values estimated using only 1 Alice policy and 1 Bob policy, which requires only 10M extra training steps ($8.33\%$ of the total budget). Nevertheless, we expect the estimation variance of LoI will matter when it comes to guiding resource allocation in more complex environments or other applications that require a more accurate estimation of LoI. 


\section{Discussion}
\subsection{Summary}

In our study, we introduce the \emph{Level of Influence} (LoI) metric, a measure that quantifies the interaction intensity between agents across varied scenarios in multi-agent reinforcement learning. Our findings demonstrate that policies trained with larger population sizes exhibit improved performance when paired with unseen co-players in highly interactive scenarios. Our proposed metric can effectively predict the potential generalization improvement by having a more diverse set of co-player policy distribution during training.

Our results strongly support the strategic allocation of resources for training a tailored set of policies across diverse scenarios guided by the LoI metric. This approach consistently yields higher average performance compared to uniform allocation across different environments with distinct game properties within the constraints of a limited computation budget.

\subsection{Limitations and Future Work}
Estimating LoI with self-play policies is susceptible to high variance. LoI essentially gauges the extent to which Alice's performance is influenced by the policy distribution of Bob. We need a diverse set of policies (\textit{i.e.}, diverse Bobs) to cover its potential distribution as much as possible in computing the LoI, while neither self-play nor population-play can guarantee such diversity to exist under limited training complexity. Thus, a delicate balance emerges between the expense of LoI estimation and its accuracy. Although we demonstrate that a high variance in LoI estimation does not necessarily hinder the effectiveness of the proposed resource allocation scheme, there may exist other downstream applications of LoI that require LoI estimation with substantially reduced variance. In future work, we aim to address this problem by exploring both theoretical grounds and practical algorithms to generate diversity guaranteed self-play policies with minimal computation cost~\cite{rahman2023minimum}, so that we can accurately estimate LoI in a sample-efficient manner. 

Moreover, while LoI serves as a valuable metric for comparison within a set of scenarios, it is environment-specific. Consequently, directly comparing the numerical values of LoI between scenarios from different environments lacks meaningful interpretation. In future work, we are interested in extending this idea in a broader context of meta-learning, where cross-environment comparisons are essential. Subsequent work may include generalizing the LoI into a comprehensive metric with predefined value ranges and thresholds across diverse environments.




\begin{acks}
We are deeply grateful to Jiaxun Cui and Arrasy Rahman for their helpful discussions and support on this work. We also thank Micah Carroll for the insightful discussion at the early stage of the project.
\end{acks}



\bibliographystyle{ACM-Reference-Format} 
\bibliography{ref}

\clearpage
\appendix
\section{Training}
\label{app:training}
This section presents the hyperparameters used for training agents. We train agents on MeltingPot substrates registered in Ray RLLib. We specify one PPO policy per agent. The hyperparameters are reported in Table~\ref{tab:hyper}.

\begin{table}[t]
  \caption{Hyperparameters for RLLib PPO policies.}
  \label{tab:hyper}
  \begin{tabular}{cc}\toprule
    Setting & Value \\ \midrule
    \# rollout workers & 2\\
    rollout fragment length & 100\\
    train batch size & 1600\\
    learning rate & 5e-5\\
    \# fully connected layers & (64, 64)\\
    \# post fcnet hiddens layers & 256\\
    LSTM size & 256\\
    SGD minibatch size & 128\\
    \# SGD iteration & 30\\
    GAE lambda & 1\\
    KL coefficient & 0.2\\
    KL target & 0.01\\
    clip parameter & 0.3\\
    value function clip parameter & 10\\
    \bottomrule
  \end{tabular}
\end{table}

\section{Additional Results}
\label{app:add_results}
This section presents additional results and statistical analyses supporting the results in the main text.

\subsection{Fixed-Bobs Evaluation}
\label{app:fixed_bobs}
\begin{figure}[t]
    \centering
    \subcaptionbox{Chicken
    \label{fig:meltingpot_bar_c}}[0.49\linewidth][c]{%
        \includegraphics[width=\linewidth]{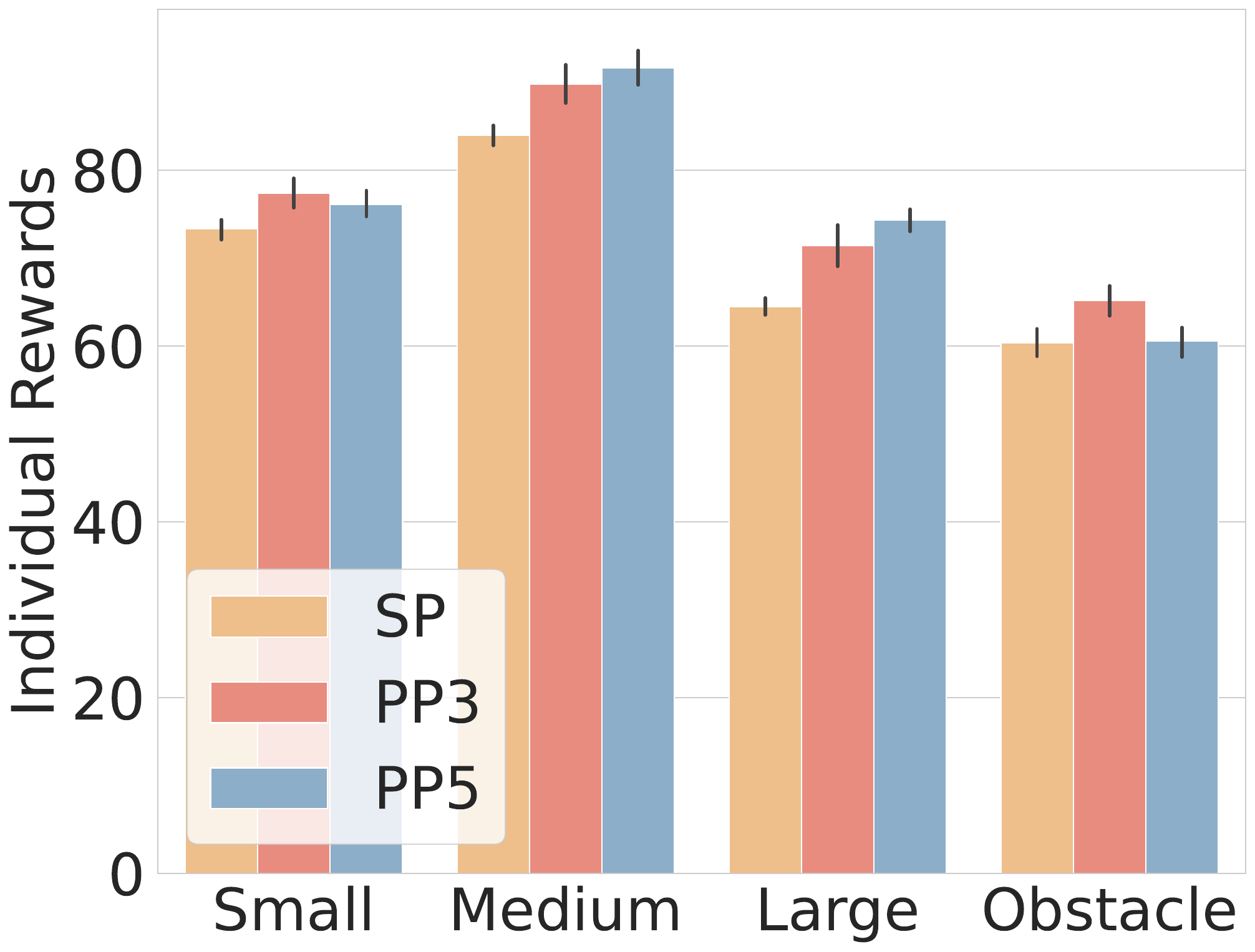}}
    \subcaptionbox{Pure Coordination
    \label{fig:meltingpot_bar_pc}}[0.49\linewidth][c]{%
        \includegraphics[width=\linewidth]{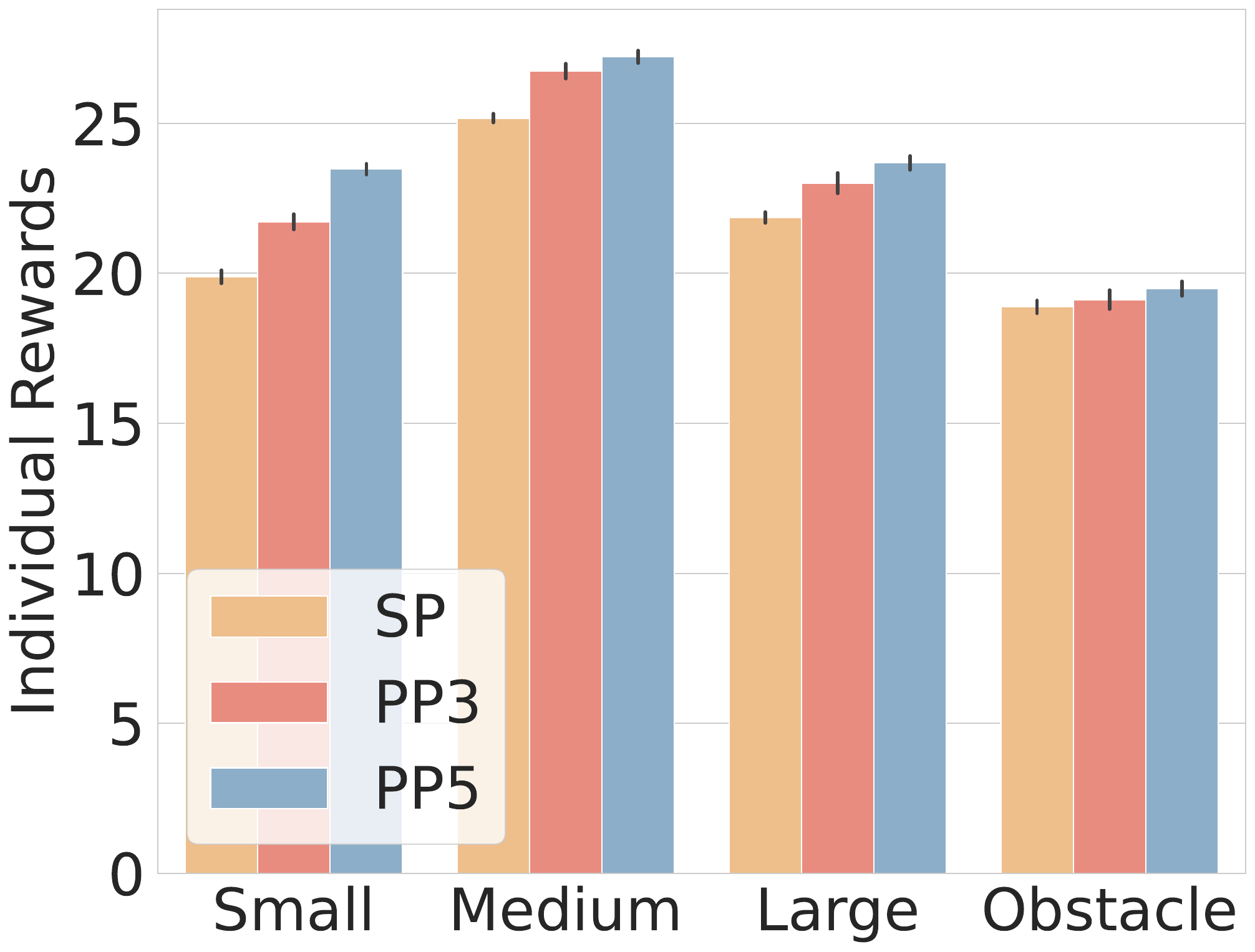}}\\
    \subcaptionbox{Prisoners Dilemma
    \label{fig:meltingpot_bar_pd}}[0.49\linewidth][c]{%
        \includegraphics[width=\linewidth]{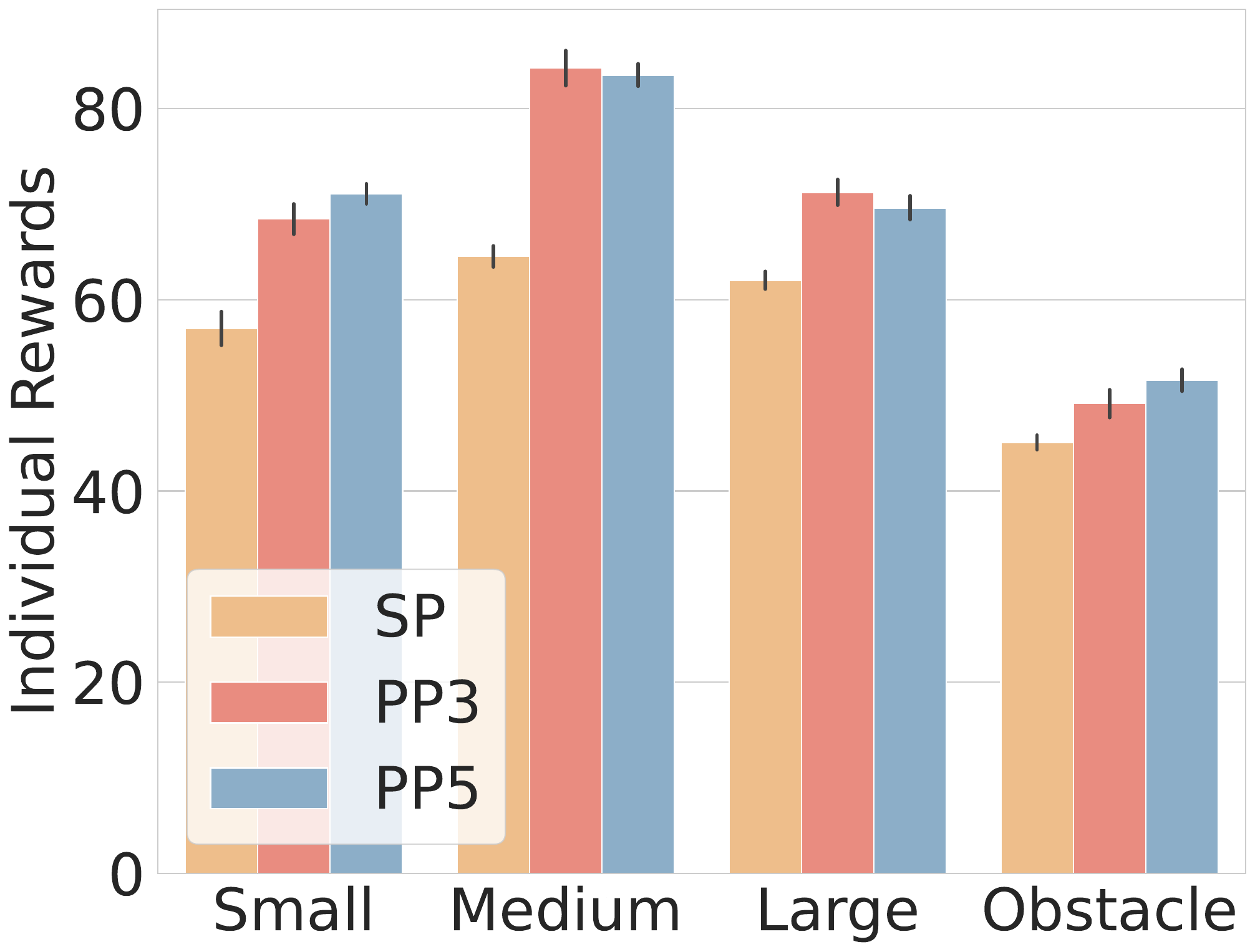}}
    \subcaptionbox{Stag Hunt
    \label{fig:meltingpot_bar_sh}}[0.49\linewidth][c]{%
        \includegraphics[width=\linewidth]{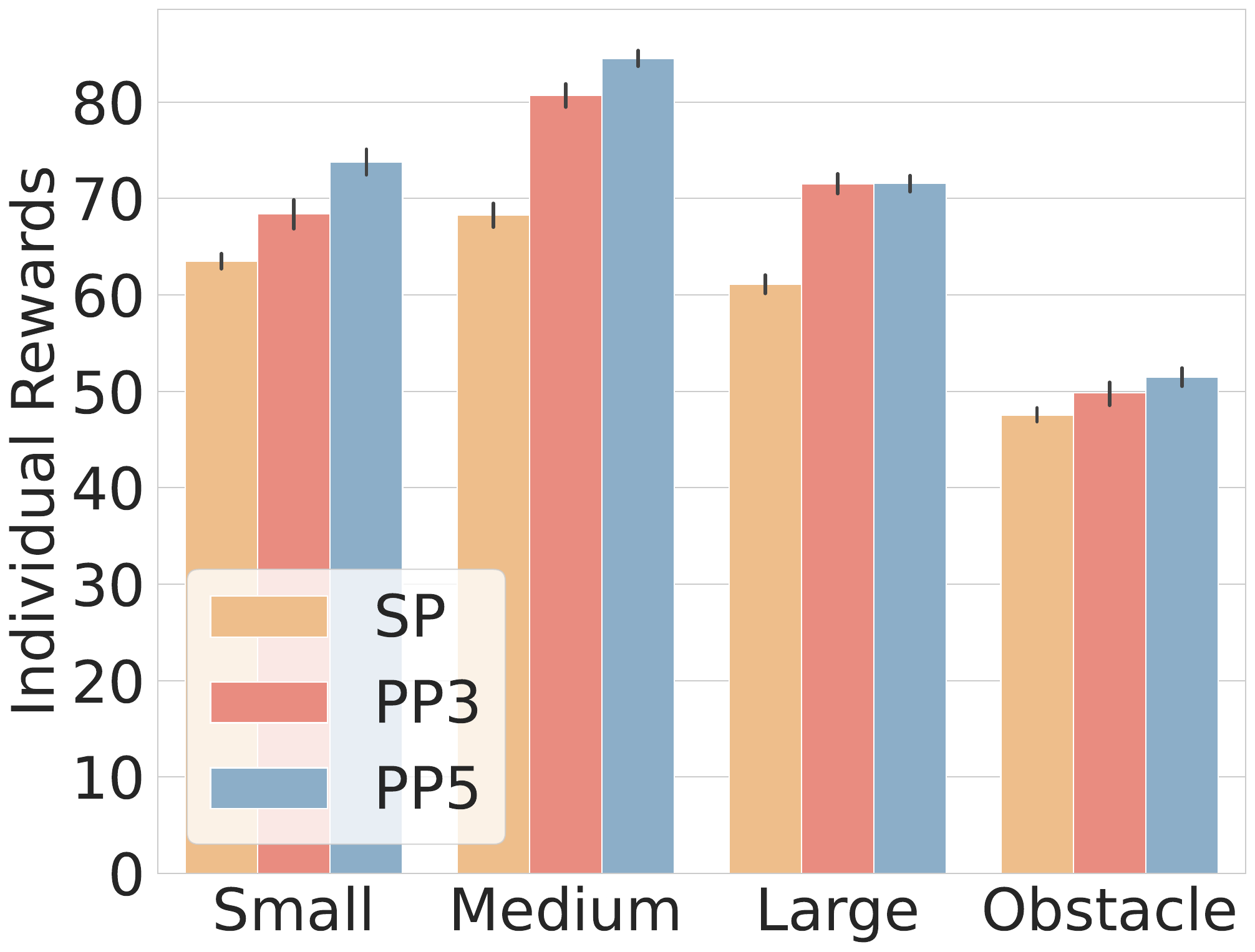}}
    \caption{Fixed-Bobs evaluation of agents trained by self-play (SP), population-play $\mathbf{p=3}$ (PP3), population-play $\mathbf{p=5}$ (PP5) across four scenarios. Error bars correspond to $\mathbf{95\%}$ confidence intervals calculated over all populations and seeds (5 seeds for SP) with 10 independent games between each Alice-Bob combination. \emph{Result:} With a growing population, PP gains a larger advantage over SP in general. However, percentage increments vary across different scenarios under the same environment, and the overall trends of improvement vary across different environments.}
    \label{fig:meltingpot_bar}
\end{figure}

A bar plot version of the Figure~\ref{fig:meltingpot_heatmap} bottom row is shown in Figure~\ref{fig:meltingpot_bar}. The numerical values are identical with additional error bars corresponding to $95\%$ confidence intervals for better comparison. Conclusions are the same as mentioned in Section~\ref{sec:FixedBob_result}.

\subsection{Resource Allocation}
\label{app:resource_alloc}
In this section, we demonstrate the effect of Bob's population number on LoI approximation and resource allocation.

\begin{figure}[!htb]
    \centering
    \subcaptionbox{Chicken
    \label{fig:index_var_c}}[0.49\linewidth][c]{%
        \includegraphics[width=\linewidth]{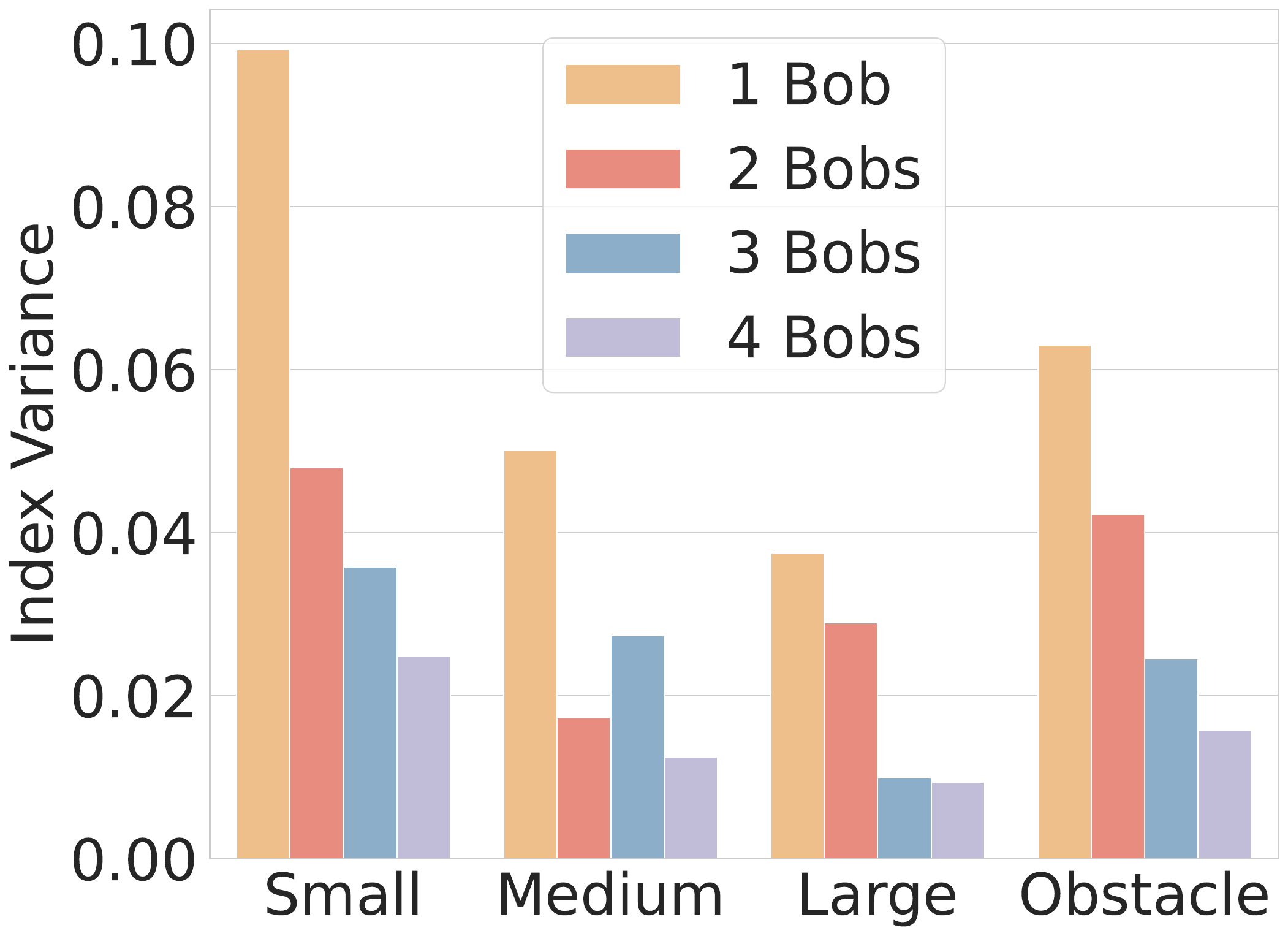}}
    \subcaptionbox{Pure Coordination
    \label{fig:index_var_pc}}[0.49\linewidth][c]{%
        \includegraphics[width=\linewidth]{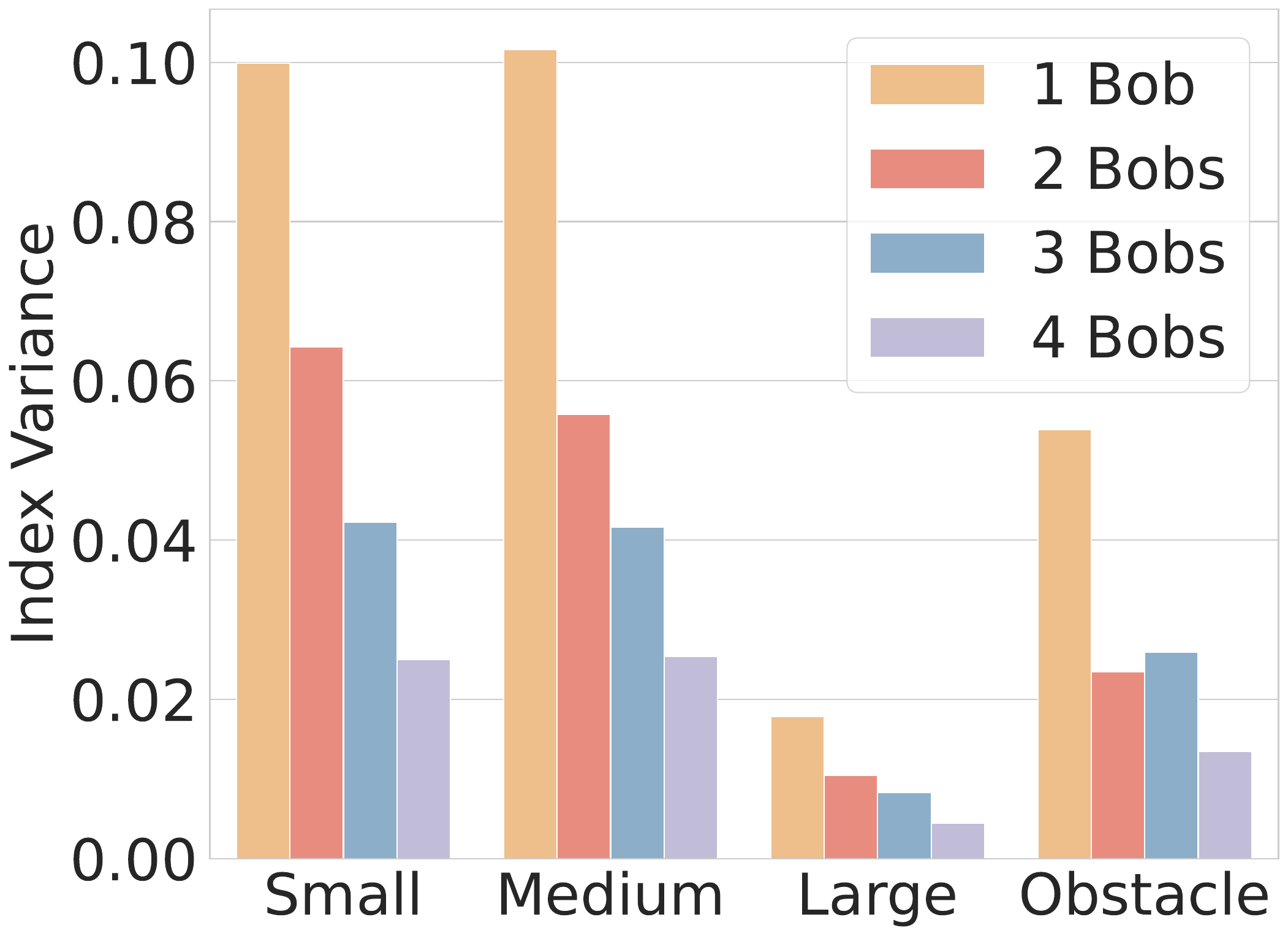}}\\
    \subcaptionbox{Prisoners Dilemma
    \label{fig:index_var_pd}}[0.49\linewidth][c]{%
        \includegraphics[width=\linewidth]{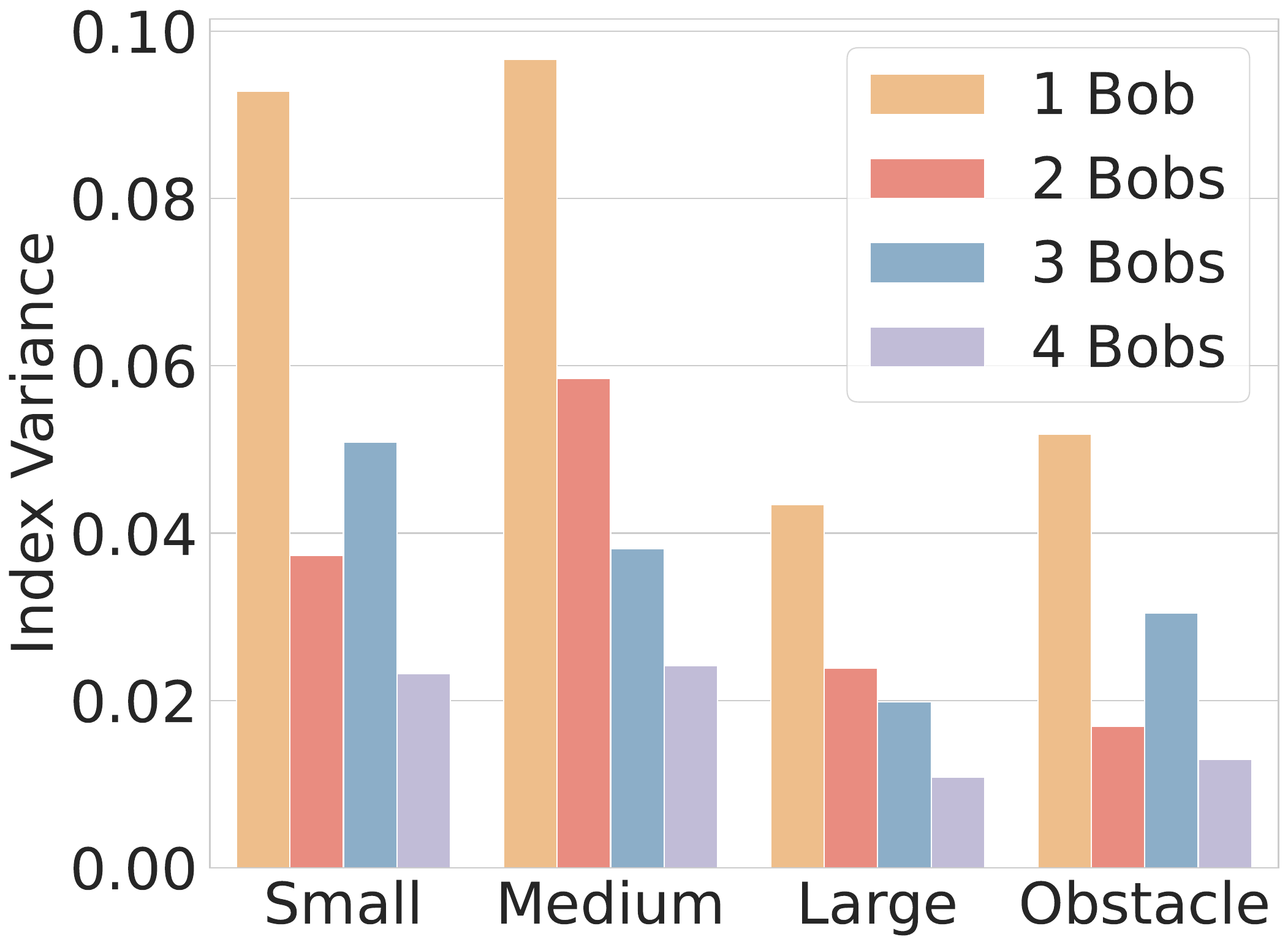}}
    \subcaptionbox{Stag Hunt
    \label{fig:index_var_sh}}[0.49\linewidth][c]{%
        \includegraphics[width=\linewidth]{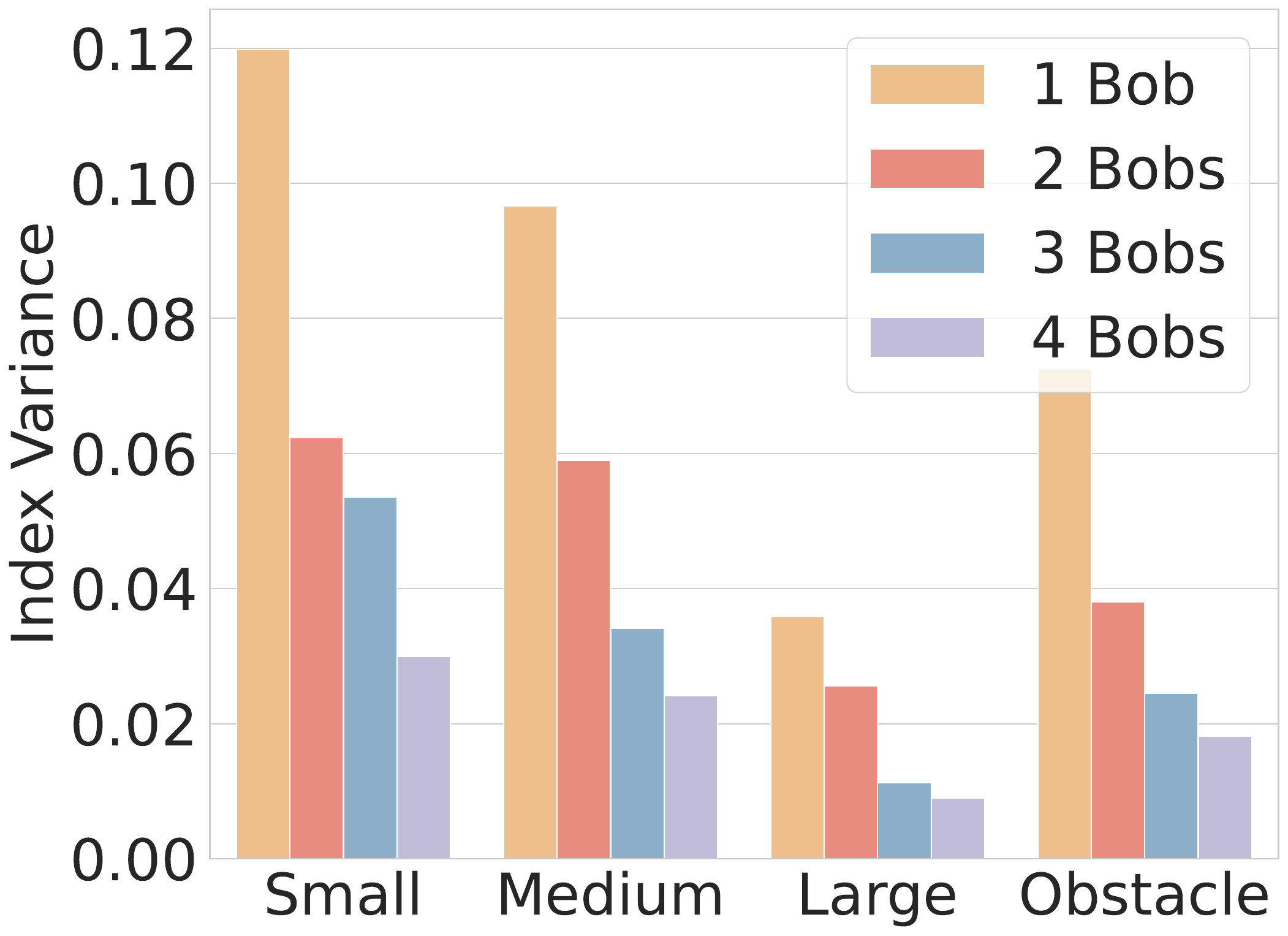}}
    \caption{Effect of Bob's population size $b$ on the variance of LoI approximation. Variance is calculated on 5 individual tests. \textit{Result:} Increasing Bob's population size lowers the LoI estimation variance across all scenarios and environments.}\label{fig:index_var}
\end{figure}

Following Section~\ref{sec:Approx} we estimate LoI by training $a+b$ SP policies with $a=1$ and $b\in\{1,2,3,4\}$. For each $b$, we train 5 independent sets of SP policies (\textit{i.e.}, 5 different sets of $a+b$ SP policies) and compute the LoIs for each set, respectively. Subsequently, We compute the variance of the estimated LoIs and report the results in Figure~\ref{fig:index_var}.

We can observe that augmenting Bob's population size leads to a notable reduction in the variance of LoI estimation. This trend persists across all scenarios and environments. However, this reduction comes at the expense of additional computational resources. 

\begin{figure}[!hbt]
    \centering
    \includegraphics[width=\linewidth]{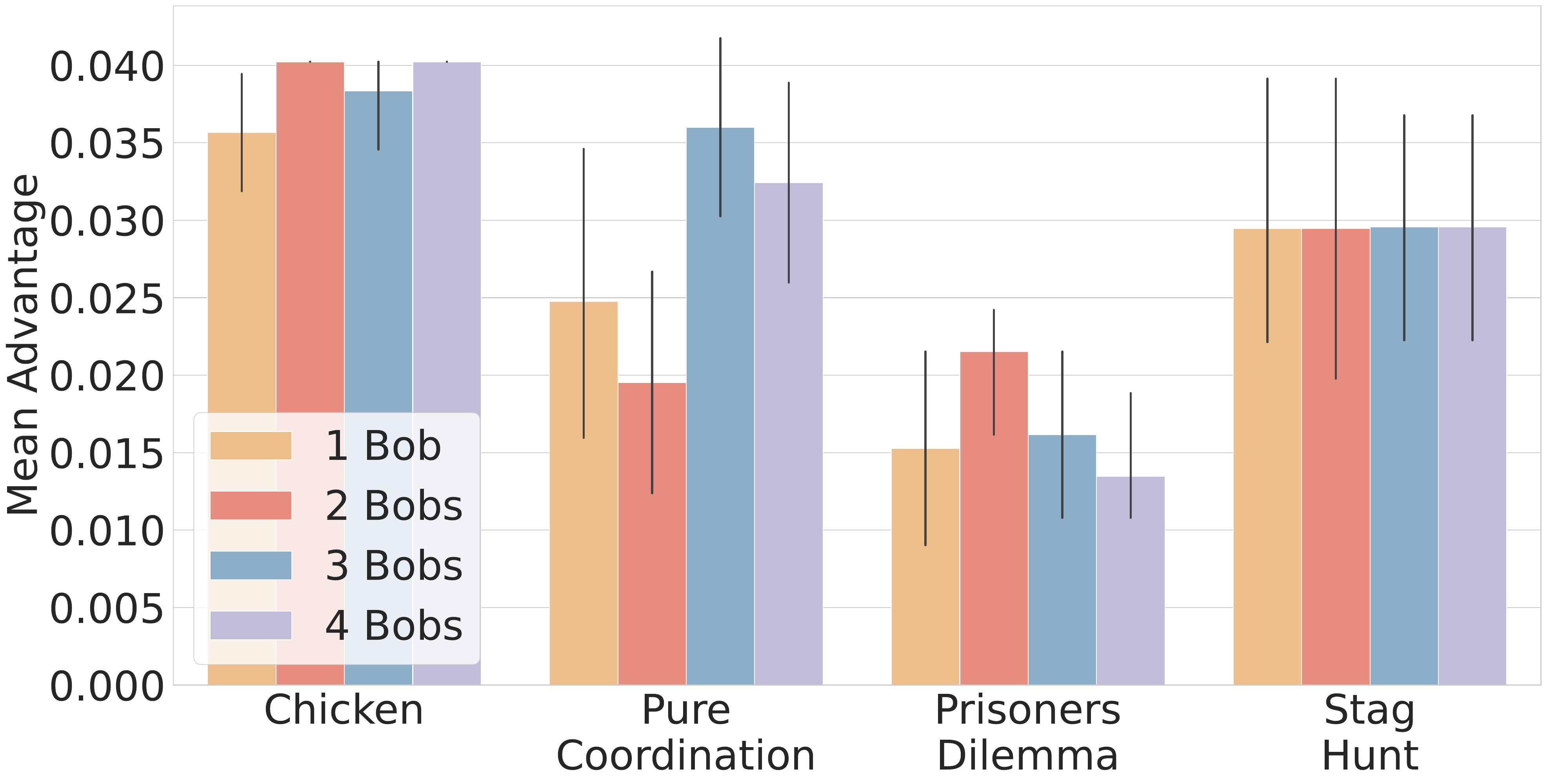}
    \caption{Effect of Bob's population size $b$ on the advantage of LoI-guided heuristic allocation over uniform allocation. Error bars correspond to $\mathbf{95\%}$ confidence intervals calculated over 5 independent comparisons. \textit{Result:} The increase in Bob's population size does not impact the advantage in the average performance of LoI-guided heuristic allocation over uniform allocation.}
    \label{fig:mean_diff}
\end{figure}

Next, we proceed with the resource allocation test outlined in Section~\ref{sec:ResouceAllo}, comparing the LoI-guided heuristic allocation and uniform allocation using the previously calculated LoI values. We define \emph{mean advantage} as the disparity in average normalized individual rewards between the heuristic and uniform cases across 5 independent experiments. The results are reported in Figure~\ref{fig:mean_diff}.

It is evident that there is not a distinct correlation between Bob's population size used in LoI calculation and the advantage of LoI-guided heuristic allocation over uniform allocation. (\textit{i.e.}, lower variance in LoI approximation does not necessarily impact the performance of heuristic allocation based on that index). As a result, a smaller population size for LoI estimation can still attain comparable performance in the resource allocation task using the proposed heuristic method.

\section{Methods}
\label{app:methods}
In this section we present the pseudocode for LoI-guided resource allocation implementation (Algorithm~\ref{alg:res_alloc}).
\begin{algorithm}
    \caption{LoI-guided resource allocation}
    \label{alg:res_alloc}
    \begin{algorithmic}[1]
        \Require \# scenarios $n$, LoIs for given scenarios $I$
        \State Get mean LoI across scenarios $\bar{I}\gets\Call{Mean}{I}$
        \State $\sigma\gets\Call{STD}{I}$
        \State $I_\mathrm{l}\gets\bar{I}-\sigma$
        \State $I_\mathrm{u}\gets\bar{I}+\sigma$
        \State Initialize set of training method $\mathcal{G}\gets\{\}$
        \State Initialize count $c\gets 0$
        \For{i=1:n}
            \If{$I_i<I_\mathrm{l}$}
                \State $\mathcal{G}\gets\mathcal{G}\cup\mathtt{'SP'}$
                \State $c\gets c+1$
            \ElsIf{$I_i>I_\mathrm{u}$}
                \State $\mathcal{G}\gets\mathcal{G}\cup\mathtt{'PP5'}$
                \State $c\gets c-1$
            \Else
                \State $\mathcal{G}\gets\mathcal{G}\cup\mathtt{'PP3'}$
            \EndIf
        \EndFor
        \If{$c\neq 0$}\Comment{Adjust methods to keep the total budget}  
            \If{$c > 0$}\Comment{Need more \texttt{'PP5'}}
                \State $k\gets\Call{ArgMax}{\mathcal{G}}$
                \State $\mathcal{G}_k\gets \texttt{'PP5'}$
            \Else\Comment{Need more \texttt{'SP'}}
                \State $k\gets\Call{ArgMin}{\mathcal{G}}$
                \State $\mathcal{G}_k\gets \texttt{'SP'}$
            \EndIf
        \EndIf
        \Ensure $\mathcal{G}$
    \end{algorithmic}
\end{algorithm}

\end{document}